\documentclass[aps,prb,twocolumn,amsmath,amssymb,nofootinbib,superscriptaddress,floatfix]{revtex4-1}
\usepackage{amsmath}
\usepackage{amssymb}
\usepackage{amsthm}
\usepackage{graphicx}

\makeatletter

\newcommand{\cblue}[1]{\textcolor{black}{#1}}
\newcommand{\cred}[1]{\textcolor{red}{#1}}

\usepackage[all,cmtip]{xy}

\begin{document}

\title{
Field theory representation of 
gauge-gravity
symmetry-protected\\
topological invariants, group cohomology and beyond 
}
\author{Juven C. Wang} \email{juven@mit.edu} 
\affiliation{Department of Physics, Massachusetts Institute of Technology, Cambridge, MA 02139, USA}
\affiliation{Perimeter Institute for Theoretical Physics, Waterloo, ON, N2L 2Y5, Canada}

\author{Zheng-Cheng Gu} \email{zgu@perimeterinstitute.ca}
\affiliation{Perimeter Institute for Theoretical Physics, Waterloo, ON, N2L 2Y5, Canada}

\author{Xiao-Gang Wen} \email{xwen@perimeterinstitute.ca}
\affiliation{Perimeter Institute for Theoretical Physics, Waterloo, ON, N2L 2Y5, Canada}
\affiliation{Department of Physics, Massachusetts Institute of Technology, Cambridge, MA 02139, USA}


\begin{abstract}
{The challenge of identifying symmetry-protected topological states (SPTs)
is due to their lack of symmetry-breaking order parameters and intrinsic topological orders.
For this reason, it is impossible to formulate SPTs under Ginzburg-Landau theory or
probe SPTs via fractionalized bulk excitations and topology-dependent ground state degeneracy.} 
However, the partition functions from path integrals with various symmetry twists
are the universal SPT invariants defining topological probe responses, 
fully characterizing SPTs.
In this work, we  use gauge fields to represent those symmetry twists in
 closed spacetimes of any dimensionality and arbitrary topology. This allows us
to express the  SPT invariants in terms of continuum field theory.
We show that SPT invariants of pure gauge actions describe the SPTs
predicted by group cohomology, while the mixed gauge-gravity actions describe
the beyond-group-cohomology SPTs. 
We find new examples of mixed gauge-gravity actions for U(1) SPTs in 
4+1D via 
mixing the gauge first Chern class with a gravitational Chern-Simons term, or viewed as a 5+1D Wess-Zumino-Witten term with a Pontryagin class. 
We rule out U(1) SPTs in 3+1D mixed with a Stiefel-Whitney class.
We also apply our approach to the bosonic/fermionic topological insulators protected by U(1) charge and $\Z_2^T$ time-reversal symmetries whose pure gauge action 
is the axion $\theta$-term.
Field theory representations of SPT invariants not only serve as tools for classifying SPTs, but also guide us in designing physical probes for them.
In addition, our field theory representations are independently powerful for studying group cohomology within the mathematical context.

\end{abstract}

\maketitle


\se{Introduction}--
Gapped systems 
without symmetry breaking\cite{GL5064,LanL58} can have intrinsic 
topological order.\cite{{Wtop},{WNtop},{Wrig}}
However, even without symmetry breaking and without topological order, gapped
systems can still be nontrivial if there is certain global symmetry
protection,
known as Symmetry-Protected Topological
states (SPTs).\cite{{Gu:2009dr},Chen:2011pg,2013arXiv1301.0861C,{Pollmann:2009}}
Their non-trivialness can be found in the gapless/topological
boundary modes protected by a global symmetry, which shows {gauge or
gravitational
anomalies}.\cite{Wen:2013oza,{Qi:2008ew},RyuZhang,Wang:2014tia,Kapustin:2014lwa,Kapustin:2014zva,{Freed:2014eja},{Cho:2014jfa},{Sule:2013qla},{Hsieh:2014lba},{Wang:2013yta},{2014arXiv1403.6491C},{Vishwanath:2012tq},{Kravec:2013pua},{Metlitski:2013uqa},K1467,K1459,{Fidkowski:2013jua},CWang1,{CWang2},KW14}
More
precisely, they are short-range entangled states which can be deformed to a
trivial product state by local unitary
transformation\cite{{Chen:2010gda},VCL0501,V0705} if the deformation breaks the
global symmetry.  Examples of SPTs are Haldane spin-1 chain protected by
spin rotational symmetry\cite{H8364,{AKL8877}} and the topological
insulators\cite{{TI4},{TI5},{TI6}} protected by fermion number conservation and
time reversal symmetry.

While some classes of topological orders can be described by topological quantum field theories (TQFT),\cite{{Schwarz:1978cn},Witten:1988hf,{Witten:1988ze},{gaugesym}}
it is less clear
{\it how to systematically construct field theory 
with a global symmetry to classify or characterize SPTs for any dimension}.
This challenge originates from the fact that SPTs 
is naturally defined on a discretized spatial lattice
or on a discretized spacetime path integral
by a group cohomology construction\cite{Chen:2011pg,{Dijkgraaf:1989pz}}
instead of continuous fields.
\cblue{Group cohomology construction of SPTs also reveals a duality
between some SPTs and the Dijkgraaf-Witten topological gauge theory.\cite{{Dijkgraaf:1989pz},{LG1209}}}

Some important progresses have been recently made to tackle the above question.
For example, there are 2+1D\cite{dim} Chern-Simons theory,\cite{{Levin:2009},{Levin:2012ta},Lu:2012dt,{Chenggu},{Ye:2013upa}}
non-linear sigma models,\cite{{LiuWen},Bi:2013oza}
and an orbifolding approach
implementing modular invariance on 1D edge modes.\cite{Sule:2013qla,{Hsieh:2014lba}}
The above approaches have their own benefits, but they may be either limited to certain dimensions, or
be limited to some special cases.
Thus, the previous works may not
fulfill all SPTs predicted from group cohomology classifications.

In this work, we will provide a more systematic way to tackle this problem, 
by constructing topological response field theory and topological invariants for SPTs (SPT invariants) in any dimension protected by a symmetry group $G$.
The new ingredient of our work suggests 
a one-to-one 
correspondence between the continuous semi-classical probe-field partition function and the discretized
cocycle of cohomology group, $\cH^{d+1}(G,\R/\Z)$, predicted to classify $d+1$D SPTs with a symmetry group $G$.\cite{refAppendix}
Moreover, our
formalism can even attain SPTs beyond group cohomology classifications.\cite{Vishwanath:2012tq,K1467,K1459,{Fidkowski:2013jua},CWang1,{CWang2}} 

\begin{figure}[!t]
\begin{center}
\includegraphics[scale=0.4]{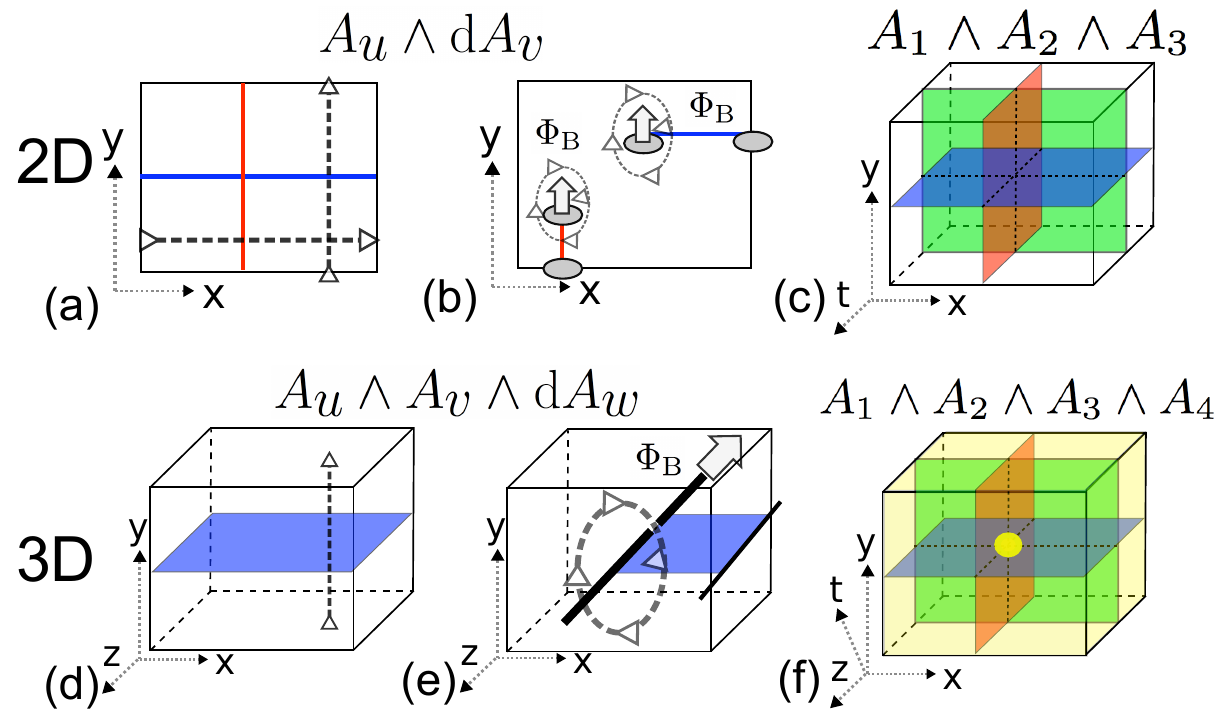} \end{center}
\caption{
On a spacetime manifold, the 1-form probe-field $A$ can be implemented on a
codimension-1 symmetry-twist\cite{{Wen:2013ue},Hung:2013cda} (with flat $\dd A=0$)
modifying the Hamiltonian $H$, but the global symmetry $G$ is preserved as a
whole. The symmetry-twist is analogous to a branch cut, going along the arrow -
- -$\vartriangleright$ would obtain an Aharonov-Bohm phase $\ep^{ig}$ with $g \in
G$ by crossing the branch cut (Fig.(a) for 2D, Fig.(d) for 3D).
{
However if the
symmetry twist ends, 
its ends 
are \emph{monodromy defects}
with $\dd A \neq 0$, effectively with a gauge flux insertion.
Monodromy
defects in Fig.(b) of 2D act like 0D point particles carrying
flux,\cite{{Wen:2013ue},{Wang:2014tia},{LG1209},{Santos:2013uda},{Barkeshli:2013yta}}
in Fig.(e) of 3D act like 1D line strings carrying flux.\cite{Wang:2014oya,{JMR1462},{WL1437},{Jian:2014vfa}}
}
The non-flat monodromy defects with $\dd A\neq 0$ are essential to realize $\int A_u \dd A_v$
and $\int A_u A_v \dd A_w$ for 2D and 3D, while the flat connections ($\dd A=0$) are
enough to realize the {\it top} Type $\int A_1  A_2  \dots A_{d+1}$ whose
partition function on a spacetime $\mathbb{T}^{d+1}$ torus with $(d+1)$
codimension-1 sheets intersection (shown in Fig.(c),(f) in 2+1D, 3+1D) renders a nontrivial element for
Eq.(\ref{eq:SPTZ}).
} \label{fig:1} \end{figure}

\se{Partition function and SPT invariants}--
For systems that realize topological orders, we can adiabatically deform the ground state $| \Psi_{g.s.} (g) \rangle$ of parameters $g$ via:
\bea
&&\langle \Psi_{g.s.} (g+\delta g) | \Psi_{g.s.} (g) \rangle \simeq \dots    \mathbf{Z}_0 \dots
\eea
to detect the volume-independent universal piece of
partition function, $\mathbf{Z}_0$, which
reveals non-Abelian geometric phase of ground states.\cite{Wilczek:1984dh,{Wrig},{Zhang:2011jd},{KW9327},{2014arXiv1401.0518M},{MW14},{KW14},{nAbgauge}}
For systems that realize  SPTs, however, their fixed-point partition
functions $\mathbf{Z}_0$ always equal to 1 due to 
its unique ground state
on any closed topology. We cannot distinguish SPTs via $\mathbf{Z}_0$.
However, due to the existence of a global symmetry, we can use  $\mathbf{Z}_0$
with the {\it symmetry twist}\cite{{Wen:2013ue},Hung:2013cda,{Hung:2013qpa}} to probe the SPTs.
To define the symmetry twist, we note that the Hamiltonian
$H=\sum_x H_x$ is invariant under the global symmetry
transformation $U=\prod_\text{all sites} U_x$, namely $H=U H U^{-1}$.
If we perform the symmetry
transformation $U'=\prod_{x\in \prt R} U_x$
only near the boundary of a region $R$ (say on one side of ${\prt R}$),
the local term $H_x$ of $H$ 
will be modified: $H_x\to H_x'|_{x\text{ near } \prt R}$.
Such a change along a codimension-1 surface is called a symmetry twist, {see Fig.\ref{fig:1}(a)(d)},
which modifies $\mathbf{Z}_0$ 
to $\mathbf{Z}_0(\text{sym.twist})$.
\cblue{Just like the geometric phases of the degenerate ground states characterize topological orders,\cite{{KW14}}}
we believe that $\mathbf{Z}_0(\text{sym.twist})$, on different spacetime manifolds and for different symmetry twists, fully characterizes
SPTs.\cite{{Wen:2013ue},Hung:2013cda}

The symmetry twist is similar to gauging the on-site
symmetry\cite{LG1209,LG14} except that the symmetry twist is non-dynamical. We can use the gauge connection 1-form $A$ to
describe the corresponding symmetry twists, with probe-fields $A$ coupling to
the matter fields of the system.  So we can write\cite{refAppendix} 
\begin{align} \label{eq:SPTZ}
\mathbf{Z}_0(\text{sym.twist})
=\ep^{\ti \mathbf{S}_0(\text{sym.twist})}=\ep^{\ti \mathbf{S}_0(A)}.
\end{align}
Here $\mathbf{S}_0(A)$ is the SPT invariant that we search for. 
Eq.(\ref{eq:SPTZ}) is a partition function of classical probe fields, or a topological response theory, obtained by integrating out the matter fields of SPTs path integral.
Below we would
like to construct possible forms of $\mathbf{S}_0(A)$ based on the following principles:\cite{refAppendix}
(1) 
$\mathbf{S}_0(A)$ is independent of spacetime metrics ({\it i.e.} topological),
(2) $\mathbf{S}_0(A)$ is gauge invariant (for both large and small gauge transformations), and
(3) ``Almost flat'' connection for probe fields.

\noindent
\underline{{\bf U(1) SPTs}}-- 
Let us start with a simple example of a single global U(1) symmetry.
We can probe the system by coupling the charge fields to an external probe 1-form field $A$ (with a U(1) gauge symmetry), and integrate out the matter fields.
In 1+1D, 
we can write down a partition function by dimensional counting:
$
\mathbf{Z} _0 (\text{sym.twist})
= \exp[{\;\ti\; \frac{\theta}{2\pi} \int F }]
$ with $F \equiv \dd A$,
this is the only term allowed by U(1) gauge symmetry 
$U^{\dagger} (A-\ti \dd) U \simeq A + \dd f$ with $U=\ep^{\ti f}$.
More generally, for an even $(d+1)$D spacetime,
$\mathbf{Z} _0(\text{sym.twist})
= \exp[{\;\ti\;  \frac{\theta}{(\frac{d+1}{2})! (2\pi)^{\frac{d+1}{2}}}   \int F \wedge F \wedge \dots}]
$.
Note that $\theta$ in such an action has no level-quantization ($\theta$ can be
an arbitrary real number).  Thus this theory does {\it not} really correspond
to any nontrivial class, because any $\theta$ is smoothly connected to
$\theta=0$ which represents a trivial SPTs.

In an odd dimensional spacetime, such as 2+1D, we have Chern-Simons coupling for the probe field action
$\mathbf{Z} _0(\text{sym.twist})=$
$\exp[{\;\ti\; \frac{k}{4\pi}  \int A \wedge \dd A}]$.
More generally, for an odd $(d+1)$D,
$
\mathbf{Z} _0(\text{sym.twist})
= \exp[{\;\ti\; \frac{2\pi k}{(\frac{d+2}{2})!(2\pi)^{(d+2)/2}}  \int A \wedge F \wedge \dots}],
$
which is known to have level-quantization $k = 2p$ with $p \in \Z$ for bosons, since U(1) is compact.
We see that only \emph{quantized} topological terms
correspond to non-trivial SPTs, the allowed responses $\mathbf{S}_0(A)$ reproduces
the group cohomology description of the U(1) SPTs: an even dimensional
spacetime has no nontrivial class, while an odd dimension has a $\Z$
class.

\noindent
\underline{{\bf $\prod_u Z_{N_u}$ SPTs}}-- 
Previously the evaluation of  U(1) field on a closed loop (Wilson-loop) $\oint A_u$ can be arbitrary values, whether the loop is contractible or not, since U(1) has continuous value.
{For finite Abelian group symmetry $G=\prod_u Z_{N_u}$ SPTs,
(1) the large gauge transformation $\delta A_u$ is identified by $2\pi$ (this also applies to U(1) SPTs). 
(2) probe fields have discrete $Z_N$ gauge symmetry,}
\bea \label{eq:intA}
\oint \delta A_u= 0{\pmod{2\pi}},\;\;\; \oint A_u =\frac{2\pi n_u}{ N_u} {\pmod{2\pi}}.\;\;\;\;\;
\eea
For a non-contractible loop (such as a $S^1$ circle of a torus), $n_u$ can be a quantized integer which thus allows large gauge transformation.
For a contractible loop, due to the fact that small loop has small $\oint A_u$ but $n_u$ is discrete, $\oint A_u=0$ and $n_u=0$,
which imply the curvature $\dd A=0$, thus $A$ is {\it flat} connection locally.

\noindent
{\bf (i).} For {\bf 1+1D},
the only quantized topological term is:
$
\mathbf{Z} _0(\text{sym.twist})=\exp[{\;\ti\;  k_{\tII}\int A_1 A_2 }].
$
Here and below we omit the wedge product
$\wedge$ between gauge fields as a
conventional notation.
Such a term is {\bf gauge invariant} under transformation if we impose flat connection $\dd A_1=\dd A_2=0$,
since $\delta(A_1 A_2)= (\delta A_1) A_2+ A_1 (\delta A_2)=(\dd f_1) A_2+A_1 (\dd f_2) =-f_1 (\dd A_2)-(\dd A_1) f_2=0$. Here we have abandoned the surface term
by considering a 1+1D
closed bulk spacetime ${\cM^2}$ without boundaries.

\noindent
{\bf $\bullet$ Large gauge transformation}: The  invariance of $\mathbf{Z} _0$ under the allowed large gauge transformation via Eq.(\ref{eq:intA}) implies that
the volume-integration of
$\int \delta(A_1 A_2)$
must be invariant mod $2\pi$, namely
$\frac{(2\pi)^2 k_{\tII}}{N_1}=\frac{(2\pi)^2 k_{\tII}}{N_2} = 0{\pmod{2\pi}}$.
{This rule implies the {\bf level-quantization}.}
\noindent
{\bf $\bullet$ Flux identification}: On the other hand, when the $Z_{N_1}$ flux from $A_1$, $Z_{N_2}$ flux from $A_2$ 
are inserted as $n_1$, $n_2$ multiple units of $2\pi/N_1$, $2\pi/N_2$,
we have $k_{\tII} \int A_1  A_2
=k_{\tII} \frac{(2\pi)^2 }{ N_1 N_2} n_1 n_2$.
We see that $k_{\tII}$ and $k'_{\tII}=k_{\tII}+\frac{N_1N_2}{2\pi}$
give rise to the same partition function $\mathbf{Z} _0$. Thus
they must be identified
$(2\pi) k_{\tII} \simeq (2\pi) k_{\tII} +N_1 N_2$,
as the rule of flux identification. These two rules impose
\bea \label{eq:Z1DAA}
\mathbf{Z} _0(\text{sym.twist})=\exp[{\;\ti\;  p_{\tII}\frac{N_1 N_2 }{(2\pi) N_{12}} \int_{\cM^2} A_1 A_2 }],
\eea
with
$k_{\tII} =p_{\tII}\frac{N_1 N_2 }{(2\pi) N_{12}} $, $p_{\tII} \in \Z_{N_{12}}$.
We abbreviate the greatest common divisor (gcd) $N_{12\dots u} \equiv  \gcd(N_1,N_2, \dots, N_u)$.
Amazingly we have independently recovered the formal group cohomology classification predicted as $\cH^2(\prod_u Z_{N_u},\R/\Z)=\prod_{u<v} \Z_{N_{uv}}$.

\noindent
{\bf (ii).} For {\bf{2+1D}},
we can propose a naive $\mathbf{Z} _0(\text{sym.twist})$ by dimensional counting,
$\exp[{\;\ti\; k_{\tIII}\int A_1 A_2 A_3}]$, which is 
gauge invariant under the flat connection condition.
By the large gauge transformation and the flux identification, we find that the level $k_{\tIII}$ is quantized,\cite{refAppendix} thus
\bea \label{eq:Z2DAAA}
\mathbf{Z} _0(\text{sym.twist})
= \exp[{\;\ti\; p_{\tIII}\frac{N_1 N_2 N_3}{(2\pi)^2 N_{123}} \int_{\cM^3} A_1 A_2 A_3}],
\eea
named as Type III SPTs with a quantized level $p_{\tIII} \in \Z_{N_{123}}$.
\cblue{The terminology ``Type'' is introduced and used in Ref.\onlinecite{deWildPropitius:1996gt} and \onlinecite{Wang:2014oya}.}
As shown in Fig.\ref{fig:1}, the geometric way to understand the 1-form probe
field can be regarded as (the Poincare-dual of) { codimension}-1
sheet assigning a group element $g \in G$ by crossing the sheet
as a branch cut.  These sheets can be regarded as the {\it symmetry
twists}\cite{{Wen:2013ue},Hung:2013cda} in the SPT Hamiltonian formulation.
When three sheets ($yt$, $xt$, $xy$ planes in Fig.\ref{fig:1}(c)) with
nontrivial elements $g_j \in Z_{N_j}$ intersect at a single point of a
spacetime $\mathbb{T}^3$ torus, it produces a nontrivial 
topological invariant 
in Eq.(\ref{eq:SPTZ}) for Type III  SPTs.


There are also other types of partition functions, which require to use
the insert flux $\dd
A\neq 0$ 
only at
the \emph{monodromy defect} (i.e. at the end of branch cut,
see Fig.\ref{fig:1}(b)) to probe
them:\cite{{Wen:2013oza},{Chenggu},{Ye:2013upa},{deWildPropitius:1996gt},{Lu:2012dt},{Kapustin:2014gua}}
\bea
\label{eq:Z2DAdA}
&& \mathbf{Z}_0(\text{sym.twist})
= \exp [{\;\ti\; \frac{p }{2\pi} \int_{\cM^3} A_u \dd A_v}],
\eea
{where $u,v$ can be either the same or different gauge fields.
They are Type I, II 
actions: ${p_{\tI,1} } \int A_1 \dd A_1$, 
${p_{\tII,12} } \int A_1 \dd A_2$, etc. 
In order to have $\ep^{\;\ti\; \frac{p_{\tII}
}{2\pi} \int_{\cM^3} A_1 \dd A_2}$ invariant under the large gauge
transformation, 
$p_{\tII}$ must be
integer.  In order to have
$\ep^{\;\ti\; \frac{p_{\tI}
}{2\pi} \int_{\cM^3} A_1 \dd A_1}$
 well-defined, we separate  
$A_1=\bar A_1+A_1^F$ to the non-flat part $A_1$ and the flat part $A_1^F$.
Its partition function becomes $\ep^{{\;\ti\; \frac{p_{\tI} }{2\pi} \int_{\cM^3} A^F_1 \dd \bar A_1}}$.\cite{refAppendix}
The invariance under the large gauge transformation of $A^F_1$
requires
$p_{\tI}$ to be quantized as integers.
We can 
further derive their level-classification via Eq.(\ref{eq:intA}) and two more conditions:
\bea  \label{eq:intdvarA}
\Ointint  \dd A_v =0 {\pmod{2\pi}},\;\;\; \Ointint  \delta \dd  A_v =0.
\eea
The first means that the net sum of all monodromy-defect fluxes on the spacetime manifold must have integer units of $2\pi$.
\cblue{Physically, 
a $2\pi$ flux configuration is trivial for a discrete symmetry group $Z_{N_v}$. Therefore two SPT invariants differ by a $2\pi$ flux configuration on their monodromy-defect should be regarded as the same one.}
The second condition means that the variation of the total flux is zero.
From the above two conditions for flux identification, we find the SPT invariant Eq.(\ref{eq:Z2DAdA}) describes the
$Z_{N_1}$ SPTs $p_\tI \in \Z_{N_1} =\cH^3(Z_{N_1},\R/\Z)$ and
the $Z_{N_1}\times Z_{N_2}$ SPTs $p_\tII \in \Z_{N_{12}}\subset \cH^3(Z_{N_1}\times Z_{N_2},\R/\Z)$.\cite{refAppendix}
}

\noindent
{\bf (iii).} For {\bf 3+1D},
we derive the {\it top} Type IV partition function that is independent of spacetime metrics:
\bea \label{eq:Z3DAAAA}
\mathbf{Z} _0(\text{sym.twist})
= \exp[{\ti\frac{p_{\tIV} N_1 N_2 N_3 N_4}{(2\pi)^3 N_{1234}} \int_{\cM^4} A_1 A_2 A_3 A_4}], \;\;\;\;\;\;
\eea
where $\dd A_i=0$ to ensure gauge invariance.
{The large
gauge transformation $\delta A_i$ of Eq.(\ref{eq:intA}),
and flux identification recover
$p_\tIV \in \Z_{N_{1234}}\subset \cH^4(\prod_{i=1}^4
Z_{N_i},\R/\Z)$.
Here the 3D SPT invariant 
is analogous to 2D, when the four
codimension-1 sheets ($yzt$, $xzt$, $yzt$, $xyz$-branes 
in Fig.\ref{fig:1}(f))
with flat $A_j$ of nontrivial element $g_j
\in Z_{N_j}$ intersect at a single point
on spacetime $\mathbb{T}^4$ torus, it renders a nontrivial  
partition function  
for the Type IV SPTs.
}

Another  
response 
is for Type III 3+1D SPTs: 
\begin{align}
\label{eq:Z3DAAdA}
\mathbf{Z} _0(\text{sym.twist})
&=\exp [\ti \int_{\cM^4} \frac{p_{\tIII} N_1N_2}{(2\pi)^2 N_{12}} A_1A_2 \dd A_3]
,
\end{align}
which is gauge invariant only if $\dd A_1=\dd A_2=0$. 
Based on Eq.(\ref{eq:intA}),(\ref{eq:intdvarA}),
the invariance under the large gauge transformations requires $p_\tIII \in \Z_{N_{123}}$. 
Eq.(\ref{eq:Z3DAAdA}) describes
Type III SPTs: $p_\tIII \in \Z_{N_{123}} \subset \cH^4(\prod_{i=1}^3
Z_{N_i},\R/\Z)$.\cite{refAppendix}

Yet another
response
is for Type II 3+1D SPTs:\cite{Levin_talk,{Chen:201301}}
\begin{align}
\label{eq:Z3DAAdA3}
\mathbf{Z} _0(\text{sym.twist})
&=\exp [\ti \int_{\cM^4} \frac{p_{\tII} N_1N_2}{(2\pi)^2 N_{12}} A_1A_2 \dd A_2].
\end{align}
The above is gauge invariant only if we choose $A_1$ and $A_2$ such that $\dd
A_1=\dd A_2 \dd A_2=0$.  We denote 
$A_2=\bar A_2+A^F_2$ where
$ \bar A_2  \dd \bar A_2=0$, $\dd A^F_2=0$, $\oint \bar A_2 =0$ mod $2\pi/N_2$,
and $\oint A^F_2 =0$ mod $2\pi/N_2$.  Note that in general $\dd \bar A_2\neq
0$, 
and Eq.(\ref{eq:Z3DAAdA3}) 
becomes $\ep^{\ti \int_{\cM^4} \frac{p_{\tII}
N_1N_2}{(2\pi)^2 N_{12}} A_1A^F_2 \dd \bar A_2}$.
The invariance under the
large gauge transformations of $A_1$ and $A^F_2$ and flux identification requires
$p_\tII \in \Z_{N_{12}} =  \cH^4(\prod_{i=1}^2 Z_{N_i},\R/\Z)$ of Type II SPTs.\cite{refAppendix}
For Eq.(\ref{eq:Z3DAAdA}),(\ref{eq:Z3DAAdA3}), we have assumed the monodromy {\emph{line defect}} at $\dd A \neq 0$ is {\emph{gapped}};\cite{{WL1437},Wang:2014oya}
for {\it gapless} defects, one will need to introduce extra anomalous {\it gapless} boundary theories.

\se{SPT invariants and physical probes}--\\
{
\underline{{\it Top types:}}\cite{refAppendix} The SPT invariants can help us to design physical probes for their SPTs, 
as \cblue{observables of numerical simulations or real experiments}.
Let us consider:
$\mathbf{Z} _0(\text{sym.twist})$$=\exp[\ti p_{}\frac{ {\prod_{j=1}^{d+1} N_j} }{(2\pi)^d N_{123 \dots (d+1)}}$$\int A_1 A_2 \dots A_{d+1} ]$,
a generic top type $\prod_{j=1}^{d+1} Z_{N_j}$ SPT invariant in $(d+1)$D, and its 
observables.\\
\noindent
$\bullet$ (1). \emph{Induced charges}:
%
If we design the space to have a topology $(S^1)^d$, and add the unit symmetry
twist of the $Z_{N_1}, Z_{N_2}, \dots$, $Z_{N_d}$ to the $S^1$ in 
$d$ directions respectively: $\oint_{S^1} A_j=2\pi  /N_j$.  The SPT invariant
 implies that such a configuration will carry a $Z_{N_{d+1}}$
charge $p_{}\frac{N_{d+1}}{N_{123 \dots (d+1)}}$.}

\noindent
{$\bullet$ (2).\emph{Degenerate zero energy modes}:
We can also apply dimensional reduction to probe SPTs. 
We can design the $d$D space as $(S^1)^{d-1} \times I$, and add
the unit $Z_{N_j}$ symmetry twists along the $j$-th $S^1$
circles for $j=3,\dots,d+1$. 
This induces a 1+1D $Z_{N_1}\times Z_{N_2}$ SPT invariant
$\exp[{\;\ti\; p_{}\frac{N_{12}}{N_{123 \dots (d+1)}}\frac{N_1 N_2}{2\pi N_{12}} \int
A_1 A_2}]$ on the 1D spatial interval $I$.
The 0D boundary of the
reduced 1+1D SPTs has degenerate zero modes that form a projective
representation of $Z_{N_1}\times Z_{N_2}$ symmetry.\cite{Wang:2014tia}
For example, dimensionally reducing 3+1D SPTs Eq.(\ref{eq:Z3DAAAA}) to this 1+1D SPTs, if we break the
$Z_{N_3}$ symmetry on the  $Z_{N_4}$ monodromy defect line,  gapless
excitations on the  defect line will be gapped.  A $Z_{N_3}$ symmetry-breaking
domain wall on the gapped monodromy defect line will carry  degenerate zero modes
that form a projective representation of $Z_{N_1}\times Z_{N_2}$ symmetry.}

\noindent
{$\bullet$ (3).\emph{Gapless boundary excitations}:
For Eq.(\ref{eq:Z3DAAAA}), we design the 3D space as $S^1\times
M^2$, and add the unit $Z_{N_4}$ symmetry twists along the $S^1$ circle. Then
Eq.(\ref{eq:Z3DAAAA}) reduces to
the 2+1D $Z_{N_1}\times Z_{N_2}\times Z_{N_3}$ SPT invariant
$ \exp[{\;\ti\;
p_{\tIV}\frac{N_{123}}{N_{1234}}\frac{N_1 N_2 N_3}{2\pi N_{123}} \int A_1 A_2
A_3}]$ 
labeled by $p_{\tIV}\frac{N_{123}}{N_{1234}} \in \Z_{N_{123}} \subset \cH^3(Z_{N_1}\times
Z_{N_2}\times Z_{N_3},\R/\Z)$.  
Namely, the $Z_{N_4}$
monodromy line defect carries gapless excitations identical to the edge modes
of the 2+1D $Z_{N_1}\times Z_{N_2}\times Z_{N_3}$ SPTs if the symmetry is
not broken.\cite{Wen:2013ue}
}


\noindent
\underline{{\it Lower types:}}\cite{refAppendix} Take 3+1D  SPTs of Eq.(\ref{eq:Z3DAAdA}) as an example,
there are at least two ways to design physical probes.  First, we can design
the 3D space as $M^2\times I$, where $M^2$ 
is punctured with ${N_3}$ \emph{identical}
monodromy defects each carrying $n_3$ unit $Z_{N_3}$ flux, namely 
%
$\Ointint  \dd A_3 = 2 \pi n_3$ of Eq.(\ref{eq:intdvarA}).
Eq.(\ref{eq:Z3DAAdA}) reduces to $ \exp[{\;\ti\;
p_{\tIII}^{}n_3\frac{N_1 N_2 }{(2\pi) N_{12}} \int A_1 A_2 }] $, which again
describes a 1+1D $Z_{N_1}\times Z_{N_2}$ SPTs, labeled by
$p_{\tIII}^{}n_3$  of
Eq.(\ref{eq:Z1DAA}) in $\cH^2(Z_{N_1}\times Z_{N_2},\R/\Z)=\Z_{N_{12}}$.
{This again has 0D boundary-degenerate-zero-modes.}

Second,
we can design the 3D space as $S^1\times M^2$
and add a symmetry twist of $Z_{N_1}$ along the $S^1$: $\oint_{S^1}
A_1=2\pi n_1 /N_1$,
then the  SPT invariant Eq.(\ref{eq:Z3DAAdA}) reduces to
$\exp[{\;\ti\; \frac{p_\tIII \; n_1 N_2 }{(2\pi)  N_{12}} \int  A_2 \dd A_3}]$,
a 2+1D $Z_{N_2}\times Z_{N_3}$ SPTs labeled by $\frac{p_\tIII \; n_1 N_2
}{N_{12}}$ of Eq.(\ref{eq:Z2DAdA}). \\
\noindent
{$\bullet$ (4).\emph{Defect braiding statistics and fractional charges}:}
These $\int A\dd A$ types in
Eq.(\ref{eq:Z2DAdA}), can be detected by the nontrivial braiding statistics of
monodromy defects, such as the particle/string defects in
2D/3D.\cite{{LG1209},Chenggu,{Wang:2014oya},{WL1437},{JMR1462},{Jian:2014vfa}} Moreover,
a $Z_{N_1}$ monodromy defect line carries gapless excitations identical to the
edge of the 2+1D $Z_{N_2}\times Z_{N_3}$ SPTs. If the  gapless excitations
are gapped by $Z_{N_2}$-symmetry-breaking, its domain wall will induce
fractional quantum numbers of $Z_{N_3}$ charge,\cite{Wang:2014tia,{Lu:2013wna}}
similar to Jackiw-Rebbi\cite{Jackiw:1975fn} or
Goldstone-Wilczek\cite{Goldstone:1981kk} effect.

\noindent
\underline{\bf U$(1)^m$ SPTs}-- 
It is straightforward to apply the above results to U$(1)^m$ symmetry.
Again, we find only trivial classes for even $(d+1)$D. For odd $(d+1)$D, we can define the lower type action:
$
\mathbf{Z} _0(\text{sym.twist})
= \exp[{\;\ti\; \frac{2\pi k}{(\frac{d+2}{2})!(2\pi)^{(d+2)/2}}  \int A_u \wedge F_v \wedge \dots}].
$
Meanwhile we emphasize that the {\it top} type
action with $k\int A_1 A_2  \dots A_{d+1}$ form will be trivial for U$(1)^m$ case since its coefficient $k$ is no longer well-defined,
at $N \to \infty$ of $(Z_{N})^m$ SPTs states.
For physically relevant $2+1$D,
$k \in 2\Z$ for bosonic  SPTs.
Thus, we will have a $\Z^m \times \Z^{m(m-1)/2}$ classification for U$(1)^m$ symmetry.\cite{refAppendix}

\se{Beyond Group Cohomology and mixed gauge-gravity actions}--
We have discussed the allowed action
$\mathbf{S}_0(\text{sym.twist})$ that is described by pure gauge fields $A_j$.  We
find that its allowed SPTs
coincide with group cohomology
results.  For a curved spacetime, we have more general topological
responses that contain both gauge fields for symmetry twists and gravitational
connections $\Gamma$ for spacetime geometry. 
Such mixed gauge-gravity
topological responses will attain 
SPTs beyond
group cohomology.  The possibility was recently discussed  in Ref.\onlinecite{{K1467,K1459}}.  Here we will propose some additional new examples
for  SPTs with U(1) symmetry.

 In {\bf 4+1D}, the following SPT 
 response exists,
\bea
\mathbf{Z} _0(\text{sym.twist})
&=& \exp[{\ti \frac{k}{3} \int_{\cM^5}  F \;   \text{CS}_3(\Gamma) }] 
 \nonumber \\
&=& \exp[{\ti \frac{k}{3} \int_{\cN^6}  F  \;  \tp_1 }],\  k \in \Z \;\;\;\;\;\;\;\; 
\eea
where $ \text{CS}_3(\Gamma) $ is the gravitations Chern-Simons 3-form and $\dd(\text{CS}_3)=\tp_1$
is the first Pontryagin class.
This 
SPT response is a Wess-Zumino-Witten form with a surface $\partial \cN^6=\cM^5$.
This renders 
an extra $\Z$-class of 4+1D
U(1) SPTs beyond group cohomology.  They have the following
physical property: If we choose the 4D space to be $S^2 \times M^2$ and
\cblue{put a U(1) monopole at the center of $S^2$}: $\int_{S^2} F=2\pi$, in the large $M^2$
limit, the effective 2+1D theory on $M^2$ space is $k$ copies of E$_8$ bosonic
quantum Hall states.  
A U(1) monopole in 4D space is a 1D loop.
{By cutting $M^2$ into two separated manifolds, each with a 1D-loop boundary,
we see U(1) monopole and anti-monopole 
as these two 1D-loops, each
loop carries $k$ copies of E$_8$ bosonic quantum Hall edge modes.\cite{E8}
}
\cblue{Their gravitational response can be detected by thermal transport with a thermal Hall conductance,\cite{Kane_Fisher} $\kappa_{xy}=8 k\frac{\pi^2 k_B^2}{3 h}T$.}
%

%

In {\bf 3+1D}, the following topological
response
exists
\bea
\mathbf{Z} _0(\text{sym.twist})
= \exp[ \frac{\ti}{ 2 } \int_{\cM^4}  F  w_2],
\eea
where $w_j$ is the $j^\text{th}$ Stiefel-Whitney (SW) class.
Let us 
design $\cM^4$ as a complex manifold, thus $w_{2j}=c_j$ mod 2.
The first Chern class
$c_1$ of the  tangent bundle of $\cM^4$ is also the first Chern class of the
determinant line bundle of the tangent bundle of $\cM^4$.
So if we choose the U(1) symmetry twist as 
the determinate line bundle of $\cM^4$,
we can write the above as ($F=2\pi c_1$):
$
\mathbf{Z} _0(\text{sym.twist})
= \exp[ \ti \pi \int_{\cM^4}  c_1  c_1]$. 
On a 4-dimensional complex manifold, we have $\tp_1=c_1^2-2c_2$.
Since the 4-manifold  $\CP^2$ is not a spin manifold, thus $w_2\neq 0$.  From
$\int_{\CP^2} \tp_1=3$, we see that $\int_{\CP^2}  c_1  c_1 =1$ mod 2. 
So the above topological response is non-trivial, and it 
\cblue{suggests} a $\Z_2$-class
of 3+1D U(1)  SPTs beyond group cohomology.
{Although this topological response is non-trivial, however,
we do \emph{not} gain extra 3+1D U(1) SPTs beyond group
cohomology, since $\exp[ \frac{\ti}{ 2 } \int_{\cN^4}  F  w_2] 
=\exp[ \frac{\ti}{ 4 \pi } \int_{\cN^4}  F \wedge F]$ on any manifold $\cN^4$, and  
since the level of $\int F \wedge F$ of U(1)-symmetry is not quantized on any manifold.\cite{Wen:2014zga}

\se{Fermionic/Bosonic topological insulators with U(1) charge and $\Z_2^T$ time-reversal symmetries }--

In 3+1D, the fermionic topological insulator as SPTs protected by U(1) charge and $\Z_2^T$ time-reversal symmetries 
is known to have an axionic $\theta$-term response. \cite{Qi:2008ew} 
We can verify the claim by our approach. In 3+1D, although we do not have a Chern-Simons form available, 
we can use the probe 
\be
\exp[ \frac{\ti k}{ 4 \pi } \int_{\cM^4}  F \wedge F] \equiv \exp[ \frac{\ti}{ 4 \pi } \frac{\theta}{ 2 \pi } \int_{\cM^4}  F \wedge F].
\ee
The time reversal symmetry $\Z_2^T$ on $F \wedge F$ is odd, so the $\theta$ must be odd as $\theta \to -\theta$ under $\Z_2^T$ symmetry.
On a spin manifold,
the $\frac{1}{ 8 \pi^2 } \int_{\cM^4}  F \wedge F$ corresponds to an integer of instanton number,
together with our large gauge transformation and flux identification, it dictates $\theta \simeq  \theta+ 2 \pi$. 
More explicitly, we recover the familiar form
$\exp[ \frac{\ti}{ 4 \pi } \frac{\theta}{ 2 \pi }\frac{1}{4} \int_{\cM^4}  \epsilon^{\mu \nu \rho \sigma } F_{\mu \nu}  F_{ \rho \sigma} \dd^4 x]$.
If the trivial vacuum has $\theta=0$, then the 3+1D fermionic topological insulator can be probed by the $\theta=\pi$ response.

The 3+1D bosonic topological insulator has the similar $\theta$-term topological response. Except that the spin structure 
is not required for bosonic systems, and the earlier quantization becomes doubled as an even integer, thus $\theta \simeq  \theta+ 4 \pi$.
If the trivial vacuum has $\theta=0$, then the 3+1D bosonic topological insulator can be probed by the $\theta=2\pi$ response.
More topological responses of fermionic/bosonic topological insulators within or beyond group cohomology are recently discussed in Refs. \onlinecite{K1467,K1459,Wen:2014zga}.

\se{Conclusion}--
The recently-found SPTs, described by group cohomology, have SPT invariants in terms of
\emph{pure gauge actions} (whose
boundaries have \emph{pure gauge anomalies}\cite{{Wen:2013oza},{Wang:2013yta},{Wang:2014tia},{Kapustin:2014lwa},{Kapustin:2014zva}}).
We have derived the formal group cohomology results from an easily-accessible field theory set-up. 
{For beyond-group-cohomology SPT invariants, while ours of bulk-onsite-unitary symmetry
are \emph{mixed gauge-gravity actions},
those of other symmetries
(e.g. anti-unitary-symmetry time-reversal $\Z_2^T$) may be \emph{pure gravity actions}.\cite{K1459}}
{%
SPT invariants can also be obtained via cobordism
theory,\cite{K1467,K1459,{Freed:2014eja}} or via \emph{gauge-gravity actions} whose
boundaries realizing \emph{gauge-gravitational anomalies}.}
We have incorporated this idea into a field theoretic framework, which should be applicable for both bosonic and fermionic SPTs and for more exotic states awaiting future explorations.

\se{Acknowledgements}--
JW wishes to thank Edward Witten for thoughtful comments during PCTS workshop at Princeton in March.
This research is supported by NSF Grant No.
DMR-1005541, NSFC 11074140, and NSFC 11274192.
Research at Perimeter Institute is supported by the Government of Canada through Industry Canada and by the Province of Ontario through the Ministry of Economic Development \& Innovation.

\begin{widetext}
\end{widetext}

\onecolumngrid
\appendix

\begin{center}
{\bf Supplemental Material}
\end{center}

\twocolumngrid

\section{``Partition functions of Fields'' - Large Gauge Transformation and Level Quantization} \label{App:levelQuant}

In this section, we will work out the details of large gauge transformations and level-quantizations for bosonic SPTs with a finite Abelian symmetry group $G=\prod_{u} Z_{N_u}$
for 1+1D, 2+2D and 3+1D.
We will briefly comment about the level modification for fermionic SPTs, and give another example for $G=\tU(1)^m$ (a product of $m$ copies of U$(1)$ symmetry) SPTs.
This can be straightforwardly extended to any dimension.

In the main text, our formulation has been focused on the 1-form field $A_\mu$ with an 
effective probed-field partition function $\mathbf{Z} _0(\text{sym.twist})=\ep^{\ti \mathbf{S}_0(A)}$. 
Below we will also mention
2-form field $B_{\mu\nu}$, 3-form field $C_{\mu\nu \rho}$, etc. We have known that for SPTs, a lattice formulation can easily couple 1-form field to the matter via $A_\mu J^\mu$ coupling.
The main concern of relegating $B$, $C$ higher forms to the Appendix without discussing them in the main text is precisely due to that
{\it it is so far unknown how to find the string} ($\Sigma^{\mu\nu}$) {\it or membrane} ($\Sigma^{\mu\nu \rho}$){\it-like excitations in the bulk SPT lattice
and further coupling via the $B_{\mu\nu} \Sigma^{\mu\nu}$, $C_{\mu\nu \rho} \Sigma^{\mu\nu \rho}$ terms.} 
However, such a challenge may be addressed in the future,
and a field theoretic framework has no difficulty to formulate them together. Therefore here we will discuss all plausible higher forms altogether.

For $G=\prod_{u} Z_{N_u}$, due to a discrete $Z_N$ gauge symmetry, and the gauge transformation ($\delta A$, $\delta B$, etc) must be identified by $2\pi$,
we have the general rules:
\bea
&&\oint A_u =\frac{2\pi n_u}{ N_u} {\pmod{2\pi}}\\
&&\oint \delta A_u= 0{\pmod{2\pi}} 
\eea
\bea
&&\Ointint B_u =\frac{2\pi n_u}{ N_u} {\pmod{2\pi}}\\
&&\Ointint \delta B_u =0{\pmod{2\pi}} \\
&&\Ointintint C_u =\frac{2\pi n_u}{ N_u} {\pmod{2\pi}}\\
&&\Ointintint \delta C_u = 0{\pmod{2\pi}}\\
&& \;\;\;\;\;\;\;\;\dots \nonumber
\eea
Here $A$ is integrated over a closed loop, $B$ is integrated over a closed 2-surface, $C$ is integrated over a closed 3-volume, etc.
The loop integral of $A$ is performed on the normal direction of a codimension-1 sheet (see Fig.\ref{fig:1}(a)(d)).
Similarly, the 2-surface integral of $B$ is performed on the normal directions of a codimension-2 sheet,
and the 3-volume integral of $C$ is performed on the normal directions of a codimension-3 sheet, etc.
The above rules are sufficient for the actions with flat connections ($\dd A=\dd B=\dd C=0$ everywhere).

Without losing generality, we consider a spacetime with a volume size $L^{d+1}$ where $L$ is the length of one dimension (such as a $\mathbb{T}^{d+1}$ torus).
The allowed large gauge transformation implies the $A$, $B$, $C$ locally can be:
\bea
&&A_{u,\mu}=\frac{2\pi n_u \dd x_\mu}{ N_u L },\;\;  \delta A_u=\frac{2\pi m_u \dd x_\mu}{ L }, \label{eq:largeA}\\
&&B_{u,\mu\nu}=\frac{2\pi n_u \dd x_\mu \dd x_\nu}{ N_u L^2 },\;\;  \delta B_{u,\mu\nu}=\frac{2\pi m_u \dd x_\mu \dd x_\nu}{ L^2 },\;\;\;\;\;\;\;\\
&&C_{u,\mu\nu\rho}=\frac{2\pi n_u \dd x_\mu \dd x_\nu \dd x_\rho}{ N_u L^3 },  \delta C_{u,\mu\nu\rho}=\frac{2\pi m_u \dd x_\mu \dd x_\nu \dd x_\rho}{ L^3 }.\;\;\;\;\;\;\\
&& \;\;\;\;\;\;\;\;\dots \nonumber
\eea
As we discussed in the main text, for some cases, if the codimension-$n$ sheet (as a branch cut) ends, then its end points are monodromy defects with non-flat connections ($\dd A \neq0$, etc).
Those monodromy defects can be viewed as external flux insertions (see Fig.\ref{fig:1}(b)(e)).
In this Appendix we only need non-flat 1-form: $\dd A \neq0$. We can imagine several monodromy defects created on the spacetime manifold,
but certain constraints must be imposed,
\bea
&&\Ointint  \dd A_v =0 {\pmod{2\pi}},  \label{eq:sumdA} \\
&&\Ointint  \delta \dd A_v =0.  \label{eq:dvarA}
\eea
This means that the sum of inserted fluxes at monodromy defects must be a multiple of $2\pi$ fluxes.
A fractional flux is allowed on some individual monodromy defects, but overall the net sum must be nonfractional units of $2\pi$ (see Fig.\ref{fig:2}).

For mixed gauge-gravity SPTs, we have also discussed its probed field partition function in terms of 
the spin connection $\boldsymbol{\omega}$, it
is simply related to the usual Christoffel symbol $\Gamma$ via a choice of local frame (via vielbein), which occurs
in gravitational effective probed-field partition function $\mathbf{Z} _0(\text{sym.twist})=\ep^{\ti \mathbf{S}_0(A,\Gamma, \dots)}$. 

We will apply the above rules to the explicit examples below.

\begin{figure}[!h]
\begin{center}
\includegraphics[scale=1.0]{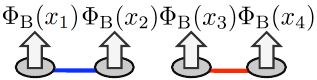} \end{center}
\caption{
The net sum of fluxes at monodromy defects (as punctures or holes of the spatial manifold) must be $2\pi n$ units of fluxes, with $n \in \Z$.
e.g. $\sum_j \Phi_\text{B}(x_j)=\iint  \dd A_v =2\pi n$.
}
\label{fig:2}
\end{figure}

\subsection{{ Top} Types: $\int A_1  A_2  \dots A_{d+1}$ with $G=\prod_{u} Z_{N_u}$}

\subsubsection{1+1D ${\int A_1 A_2 }$ }

For 1+1D bosonic SPTs with a symmetry group $G=\prod_{u} Z_{N_u}$,
by dimensional counting, one can think of $\int \dd A=\int F$, but we know that due to $F=\dd A$ is a total derivative, so it is not a bulk topological term but only a surface integral.
The only possible term is $\exp[{\;\ti\;  k_{\tII}\int A_1 \wedge A_2 }]$, (here $A_1$ and $A_2$ come from different symmetry group $Z_{N_1},Z_{N_2}$, otherwise $A_1 \wedge A_1=0$ due to
anti-symmetrized wedge product).
Below we will omit the wedge product $\wedge$ as conventional and convenient notational purposes, so $A_1 A_2 \equiv A_1 \wedge A_2$.
Such a term $A_1 A_2$ is invariant under transformation if we impose flat connection $\dd A_1=\dd A_2=0$,
since $\delta(A_1 A_2)= (\delta A_1) A_2+ A_1 (\delta A_2)=(\dd f_1) A_2+A_1 (\dd f_2) =-f_1 (\dd A_2)-(\dd A_1) f_2=0$. Here we have abandoned the surface term
if we consider a closed bulk spacetime without boundaries.\\

\noindent
{\bf $\bullet$ Large gauge transformation}: The partition function 
$\mathbf{Z} _0(\text{sym.twist})$ invariant under the allowed large gauge transformation via Eq.(\ref{eq:largeA}) implies
\bea
&&k_{\tII} \int \delta(A_1 A_2)=k_{\tII} \int (\delta A_1) A_2+A_1 (\delta A_2)\nonumber\\
&&=k_{\tII} \int \frac{2\pi m_1 \dd x_1}{  L } \frac{2\pi n_2 \dd x_2}{ N_2 L }+ \frac{2\pi n_1 \dd x_1}{ N_1 L } \frac{2\pi m_2 \dd x_2}{  L }  \nonumber\\
&&=k_{\tII}(2\pi)^2(\frac{m_1 n_2}{N_2}+\frac{n_1 m_2}{N_1} ), \nonumber
\eea
which action must be invariant mod $2\pi$ for any large gauge transformation parameter (e.g. $n_1,n_2$), namely
\bea
&&\frac{(2\pi)^2 k_{\tII}}{N_1}=\frac{(2\pi)^2 k_{\tII}}{N_2} = 0{\pmod{2\pi}} \nonumber\\
&&\Rightarrow \frac{(2\pi) k_{\tII}}{N_1}=\frac{(2\pi) k_{\tII}}{N_2} = 0{\pmod{ 1}}
\eea
\cblue{This rule of large gauge transformation implies the {\bf level-quantization}.}

\noindent
{\bf $\bullet$ Flux identification}: On the other hand, when the $Z_{N_1}$ flux from $A_1$ and $Z_{N_2}$ flux from $A_2$ are
inserted as 
$n_1$, $n_2$ multiple units of $2\pi/N_1$, $2\pi/N_2$,
we have
\bea
&&k_{\tII} \int A_1  A_2=k_{\tII} \int \frac{2\pi n_1 \dd x }{ N_1 L} \frac{2\pi n_2 \dd t}{N_2 L} \nonumber\\
&&=k_{\tII}  \frac{(2\pi)^2 }{ N_1 N_2} n_1 n_2. \nonumber
\eea
No matter what value $n_1 n_2$ is, whenever $k_{\tII}  \frac{(2\pi)^2 }{ N_1 N_2}$ shifts by $2\pi$, the symmetry-twist partition function
$\mathbf{Z} _0(\text{sym.twist})$ is invariant.
The coupling $k_{\tII}$ must be identified, via
\bea
&& (2\pi) k_{\tII} \simeq (2\pi) k_{\tII} +N_1 N_2.
\eea
($\simeq$ means the level identification.)
We call this rule as the flux identification. These two rules above imposes that $k_{\tII} =p_{\tII}\frac{N_1 N_2 }{(2\pi) N_{12}} $ with $p_{\tII}$ defined by $p_{\tII} \pmod{N_{12}}$ so
$p_{\tII} \in \Z_{N_{12}}$, where ${N_{12}}$ is the greatest common divisor(gcd) defined by $N_{12\dots u} \equiv  \gcd(N_1,N_2, \dots, N_u)$.
$N_{12}$ is the largest number can divide $N_1$ and $N_2$ from {\it Chinese remainder theorem}. We thus derive
\bea
\mathbf{Z} _0(\text{sym.twist})
= \exp[{\;\ti\; p_{\tII}\frac{N_1 N_2 }{(2\pi) N_{12}} \int_{\cM^2} A_1 A_2 }]. \;\;\;\;
\eea

\subsubsection{2+1D ${\int A_1 A_2 A_3}$ }

In 2+1D, we have $\exp[{\;\ti\;  k_{\tIII}\int A_1 A_2 A_3}]$ allowed by flat connections. We have 
the two rules, {\bf large gauge transformation}
\bea
&&k_{\tIII} \int \delta(A_1 A_2 A_3) \nonumber\\
&&=k_{\tIII} \int (\delta A_1) A_2 A_3+A_1 (\delta A_2) A_3+A_1 A_2 (\delta A_3) \nonumber\\
&&=k_{\tIII}(2\pi)^3(\frac{m_1 n_2 n_3}{N_2 N_3}+\frac{n_1 m_2 n_3}{N_1 N_3}+\frac{n_1 n_2 m_3}{N_1 N_2} ), \nonumber
\eea
which action must be invariant mod $2\pi$ for any large gauge transformation parameter (e.g. $n_1,n_2 ,\dots$)
and {\bf flux identification} with
$k_{\tIII} \int A_1  A_2 A_3= k_{\tIII} \int \frac{2\pi n_1 \dd x }{ N_1 L} \frac{2\pi n_2 \dd y}{N_2 L}  \frac{2\pi n_3 \dd t}{N_3 L} =k_{\tIII}  \frac{(2\pi)^3 }{ N_1 N_2 N_3} n_1 n_2 n_3$.
Both  large gauge transformation and flux identification respectively impose
\bea
&&\frac{(2\pi)^2 k_{\tIII}}{N_u N_v} = 0{\pmod{1}},\\
&& (2\pi)^2 k_{\tIII} \simeq (2\pi)^2 k_{\tIII} +N_1 N_2 N_3,
\eea
with $u,v \in \{ 1,2,3\}$ and $u \neq v$.
We thus derive $ k_{\tIII}=p_{\tIII}\frac{N_1 N_2 N_3}{(2\pi)^2 N_{123}} $ and
\bea
\mathbf{Z} _0(\text{sym.twist})
= \exp[{\;\ti\; p_{\tIII}\frac{N_1 N_2 N_3}{(2\pi)^2 N_{123}}  \int_{\cM^{3}} A_1 A_2 A_3}],\;\;\;\;\;\;\;\;\;\;
\eea
with $p_{\tIII}$ defined by $p_{\tIII} \pmod{N_{123}}$, so $p_{\tIII} \in \Z_{N_{123}}$.

\subsubsection{$(d+1)$D ${\int A_1 A_2 \dots A_{d+1}}$ }

In $(d+1)$D, similarly, we have $\exp[{\;\ti\;  k_{}\int A_1 A_2 \dots A_{d+1} }]$ allowed by flat connections, where
the {\bf large gauge transformation} and {\bf flux identification} respectively constrain
\bea
&&\frac{(2\pi)^d \; k_{} \; N_{u} }{\prod_{j=1}^{d+1} N_j} = 0{\pmod{1}},\\
&& (2\pi)^d k_{} \simeq (2\pi)^d k_{} +{\prod_{j=1}^{d+1} N_j},
\eea
with $u \in \{ 1,2,\dots, d+1\}$.
We thus derive
\be
\mathbf{Z} _0(\text{sym.twist})
= \exp[{\;\ti\; p_{}\frac{ {\prod_{j=1}^{d+1} N_j} }{(2\pi)^d N_{123 \dots (d+1)}} \int A_1 A_2 \dots A_{d+1} }],\;\;\;\;\;\;\;
\ee
with $p_{}$ defined by $p_{} \pmod{N_{123 \dots (d+1)}}$. We name this form
$\int A_1 A_2 \dots A_{d+1}$ as the {\bf Top Types}, which can be realized for
all flat connection of $A$.  Its path integral interpretation is a direct
generalization of Fig.\ref{fig:1}(c)(f), when the $(d+1)$ number of codimension-1 sheets
with flat $A$ on $\mathbb{T}^{d+1}$ spacetime torus with nontrivial elements
$g_j \in Z_{N_j}$ intersect at a single point, it renders a nontrivial
partition function of Eq.(\ref{eq:SPTZ}) with $\mathbf{Z} _0(\text{sym.twist}) \neq 1$.

\subsection{Lower Types in 2+1D with $G=\prod_{u} Z_{N_u}$}

\subsubsection{ ${\int A_u \dd A_v}$} \label{sec:AdA}

Apart from the top Type, we also have $\mathbf{Z} _0(\text{sym.twist})=
\exp[{\;\ti\; k\int A_u \dd A_v}]$ assuming that $A$ is almost flat but $\dd A \neq
0$ at monodromy defects.  Note that $\dd A$ is the flux of the monodromy
defect, which is an external input and does not have any dynamical variation, $\delta (\dd
A_v)=0$ as Eq.(\ref{eq:dvarA}).  For the {\bf large gauge transformation}, we
have $k_{}\int \delta(A_u \dd A_v)$ as
\bea
&&k_{}\int \big((\delta A_u) \dd A_v +A_u \delta(\dd A_v)\big)= 0 {\pmod{2\pi}} \;\;\;\; \nonumber\\
&&\Rightarrow \frac{ k_{} }{ 2\pi}\int \big( \frac{2\pi m_u \dd x }{ L} \frac{2\pi n_v \dd y\dd t}{L^2} +0  \big)=0 \pmod 1, \nonumber
\eea
for any $m_u,n_v$. We thus have
\be
 {(2\pi) k_{}}  =0 \pmod 1.
\ee
The above include both Type I and Type II  SPTs in 2+1D:
\bea
&&\mathbf{Z} _0(\text{sym.twist})
= \exp[{\;\ti\; \frac{p_{\tI} }{(2\pi) } \int_{\cM^3} A_1 \dd A_1}],  \label{eq:Z2DAdA1}\\  
&&\mathbf{Z} _0(\text{sym.twist})
= \exp[{\;\ti\; \frac{p_{\tII} }{(2\pi) } \int_{\cM^3} A_1 \dd A_2}], \label{eq:Z2DAdA2}   
\eea
where $p_{\tI}$, $p_{\tII} \in \Z$ integers.\\

\noindent
{{\bf Configuration}:
In order for Eq.(\ref{eq:Z2DAdA2}), $\ep^{\;\ti\; \frac{p_{\tII}
}{2\pi} \int_{\cM^3} A_1 \dd A_2}$ to be invariant under the large gauge
transformation that changes $\oint A_1 $ by $2\pi$, $p_{\tII}$ must be
integer.  In order for Eq.(\ref{eq:Z2DAdA1}) 
to be well defined, we denote 
$A_1=\bar A_1+A_1^F$
where $ \bar A_1  \dd \bar A_1=0$, $\dd A^F_1=0$, $\oint \bar A_1 =0$
mod $2\pi/N_1$, and $\oint A^F_1 =0$ mod $2\pi/N_1$.  In this case
Eq.(\ref{eq:Z2DAdA1}) 
becomes
$\ep^{{\;\ti\; \frac{p_{\tI} }{2\pi} \int_{\cM^3} A^F_1 \dd \bar A_1}}$.
The invariance under the large gauge transformation of $A^F_1$ requires
$p_{\tI}$ to be quantized as integers.
}

For the {\bf flux identification}, we compute $k\int A_u \dd A_v=k \int \frac{2\pi n_u d x }{ N_u L} \frac{2\pi n_v dy dt}{L^2} =k  \frac{(2\pi)^2 }{ N_u} n_u n_v$,
where $k_{}$ is identified by
\bea
&& (2\pi) k_{} \simeq (2\pi) k_{} +N_{u}.
\eea
On the other hand, the integration by parts in the case on a closed (compact without boundaries) manifold implies another condition,
\bea
&& (2\pi) k_{} \simeq (2\pi) k_{} +N_{v},
\eea


{
{\bf Flux identification}:
If we view $k_{} \simeq  k_{} +N_{u}/(2\pi)$ and $k_{} \simeq  k_{} +N_{v}/(2\pi)$ as the identification of level $k$, then
we should search for the smallest period from their linear combination.
From {\it Chinese remainder theorem}, overall the linear combination $N_{u}$ and $N_{v}$ provides the smallest unit as their greatest common divisor(gcd) $N_{uv}$:
\be
(2\pi) k_{} \simeq (2\pi) k_{} +N_{uv}
\ee
Hence $p_{\tI}$, $p_{\tII}$ are defined as $p_{\tI}{\pmod{N_1}}$ and $p_{\tII}{\pmod{N_{12}}}$,
so it suggests that $p_{\tI} \in \Z_{N_{1}}$ and $p_{\tII} \in \Z_{N_{12}}$.
}


Alternatively, using 
the fully-gauged {braiding statistics approach among particles,\cite{{Chenggu},{Ye:2013upa}}
it also renders $p_{\tI} \in \Z_{N_{1}}$ and $p_{\tII} \in \Z_{N_{12}}$.

\subsubsection{ ${\int A_1 B_2}$} \label{sec:A1B2}
For $A_u \dd A_v$ action, we have to introduce non-flat $\dd A \neq 0$ at some monodromy defect.
There is another way instead to formulate it by introducing flat 2-form $B$ with $\dd B=0$. The partition function 
$\mathbf{Z} _0(\text{sym.twist})= \exp[{\;\ti\; k_{\tII}\int A_1 B_2}]$. The {\bf large gauge transformation} and the {\bf flux identification} constrain respectively
\bea
&&\frac{(2\pi) k_{\tII}}{N_u } = 0{\pmod{1}},\\
&& (2\pi) k_{\tII} \simeq (2\pi) k_{\tII} +N_1 N_2,
\eea
with $u \in \{ 1,2\}$. We thus derive
\bea
&&\mathbf{Z} _0(\text{sym.twist})
= \exp[{\;\ti\; p_{\tII}\frac{N_1 N_2}{(2\pi) N_{12}} \int_{\cM^{3}} A_1 B_2}],\;\;\;\;
\eea
with $p_{\tII}$ defined by $p_{\tII} \pmod{N_{12}}$ and $p_{\tII} \in \Z_{N_{12}}$.

\subsection{Lower Types in 3+1D with $G=\prod_{u} Z_{N_u}$}

\subsubsection{ ${\int A_u A_v \dd A_w}$}


{
To derive $\int A_u \wedge A_v \wedge \dd A_w$ topological term, 
we first know that the $\int F_u\wedge F_v=\int \dd A_u\wedge \dd A_v$ term is only a trivial surface term 
for the symmetry group $G=\prod_j Z_{N_j}$ and for $G=\tU(1)^m$.
First, 
the flat connection $\dd A=0$
imposes that $F_u\wedge F_v=0$. 
Second, for a nearly flat connection $\dd A \neq 0$,
we have $\frac{k}{2\pi} \dd A_u\wedge \dd A_v \neq 0$ but the level quantization imposes $k \in \Z$,
and the flux identification ensures that 
$k \simeq k+1$. So all $k \in \Z$ is identical to the trivial class $k=0$.
Hence, for $G=\prod_j Z_{N_j}$, the only lower type of SPTs we have is that $\int A_u A_v \dd A_w$.
Such term vanishes for a single cycle group ($A_1 A_1 \dd A_1=0$ for $G=Z_{N_1}$, since $A_1 \wedge A_1=0$)  
thus it must come from two or three cyclic products ($Z_{N_1} \times Z_{N_2}$ or $Z_{N_1} \times Z_{N_2} \times Z_{N_3}$).
}

{\bf 3+1D bosonic topological insulator}:
However, we should remind the reader that if one consider a different symmetry group, such as 
$G=\tU(1) \rtimes Z_2^T$ of a bosonic topological insulator,
the extra time reversal symmetry $Z_2^T$ can distinguish 
two distinct classes of $\theta=0$ and $\theta=2\pi$ 
for the probe-field partition function
\be 
\exp[ \frac{\ti}{ 4 \pi } \frac{\theta}{ 2 \pi } \int_{\cM^4}  F \wedge F].
\ee
The time reversal symmetry $\Z_2^T$ on $F \wedge F$ is odd, so the $\theta$ must be odd as $\theta \to -\theta$ under $\Z_2^T$ symmetry.
The $\frac{1}{ 4 \pi^2 } \int_{\cM^4}  F \wedge F$ corresponds to an integer of instanton number,
together with our large gauge transformation and flux identification, it dictates $\theta \simeq  \theta+ 4 \pi$. 
More explicitly, we recover the familiar form
$\exp[ \frac{\ti}{ 4 \pi } \frac{\theta}{ 2 \pi }\frac{1}{4} \int_{\cM^4}  \epsilon^{\mu \nu \rho \sigma } F_{\mu \nu}  F_{ \rho \sigma} \dd^4 x]$.
If the trivial vacuum has $\theta=0$, then the 3+1D bosonic topological insulator can be probed by $\theta=2\pi$ response.

Similar to Sec.\ref{sec:AdA}, the almost flat connection but with $\dd A \neq 0$ at the monodromy defect introduces a path integral,
\bea
&& \mathbf{Z} _0(\text{sym.twist})
= \exp[{\;\ti\;  k \int_{\cM^{4}} A_u A_v \dd A_w}].
\eea
For the {\bf large gauge transformation}, we thus have
$
k_{}\int \delta(A_u A_v \dd A_w) =k_{}\int (\delta A_u)  A_v \dd A_w$ $+A_u (\delta A_v) \dd A_w$ $+A_u A_v \delta  (\dd A_w)$ $= 0 {\pmod{2\pi}}
$
$
\Rightarrow \frac{ k_{} }{ 2\pi}\int \frac{2\pi n_u \dd x }{ L} \frac{2\pi n_v \dd y }{N_v L}  \frac{2\pi n_w \dd z \dd t}{L^2} +\frac{2\pi n_u \dd x }{N_u  L} \frac{2\pi n_v \dd y }{ L}  \frac{2\pi n_w \dd z \dd t}{L^2}   =0 \pmod 1.
$
This constrains that
\be \label{eq:1}
\frac{(2\pi)^2 k_{}}{N_u} =\frac{(2\pi)^2 k_{}}{N_v}  =0 \pmod 1.
\ee
Thus, the large gauge transformation again implies that $k$ has a {\bf level quantization}.

For the {\bf flux identification},
$ k\int A_u A_v \dd A_w=k \int \frac{2\pi n_u d x }{ N_u L} \frac{2\pi n_v dy }{N_v L}  \frac{2\pi n_w dz dt}{L^2} =k  \frac{(2\pi)^3 }{ N_u N_v} n_u n_v n_w$.
The whole action is identified by $2\pi$ under the shift of quantized level $k$:
%
\be \label{eq:3+1DTypeII_Q}
(2\pi)^2 k_{} \simeq (2\pi)^2 k_{} +N_{u} N_{v}.
\ee


{
For the case of a $Z_{N_1} \times Z_{N_2}$ symmetry, we have Type II SPTs. 
We obtain a partition function:
\bea \label{eq:AppA1A2dA2}
&&\mathbf{Z} _0(\text{sym.twist})
= \exp[{\;\ti\; p_{\tII}\frac{N_1 N_2 }{(2\pi)^2 N_{12}} \int_{\cM^{4}}  A_1 A_2 \dd A_2}],\;\;\;\;\;\;\;\;\;\; 
\eea
The flux identification Eq.(\ref{eq:3+1DTypeII_Q}) implies that
the identification of $p_{\tII} \simeq p_{\tII} + {N_{12}}$.
Thus, it suggests that a cyclic period of $p_{\tII}$ is ${N_{12}}$,
and we have $p_{\tII} \in \Z_{N_{12}}$.}

{
Similarly, there are also distinct
classes of Type II SPTs with a partition function
$\exp[{\;\ti\; p_{\tII}\frac{N_1 N_2 }{(2\pi)^2 N_{12}} \int_{\cM^{4}}  A_2 A_1 \dd A_1}]$ with $p_{\tII} \in \Z_{N_{12}}$.
We notice that $A_1 A_2 \dd A_2$ and $A_2 A_1 \dd A_1$ are different types of SPTs, because they are not identified even
by doing integration by parts.} 

{
For the case of $Z_{N_1} \times Z_{N_2} \times Z_{N_3}$ symmetry, we have extra Type III SPTs partition functions (other than the above Type II SPTs), for example:
\bea \label{eq:AppA1A2dA3}
&&\mathbf{Z} _0(\text{sym.twist})
= \exp[{\;\ti\; p_{\tIII}\frac{N_1 N_2 }{(2\pi)^2 N_{12}} \int_{\cM^{4}}  A_1 A_2 \dd A_3}]. \;\;\;\;\;\;\;\;\;
\eea
Again, the flux identification Eq.(\ref{eq:3+1DTypeII_Q}) implies that
the identification of
\be \label{eq:levelA1A2dA3}
p_{\tIII} \simeq p_{\tIII} + {N_{12}}.
\ee
Thus, it suggests that a cyclic period of $p_{\tIII}$ is ${N_{12}}$,
and $p_{\tIII} \in \Z_{N_{12}}$.}

{
However, there is an extra constraint on the level identification. Now consider $ \int A_1 A_2 \dd A_3=\int - \dd ( A_1 A_2)  A_3$ up to a surface integral $\int d (A_1 A_2 A_3)$.
Notice that $\int -\dd ( A_1 A_2) \dd A_3=-\int A_2 A_3 \dd A_1 -\int A_3 A_1 \dd A_2$.
If we reconsider the flux identification of Eq.(\ref{eq:AppA1A2dA3}) in terms of
$\mathbf{Z} _0(\text{sym.twist})=
 \exp[{\;-\ti\; p_{\tIII}\frac{N_1 N_2 }{(2\pi)^2 N_{12}} \int_{\cM^{4}}   (A_2 A_3 \dd A_1+A_3 A_1 \dd A_2)}]$, we find
 the spacetime volume integration yields a phase
$\mathbf{Z} _0(\text{sym.twist}) $$=$$\exp[{\;-\ti\; p_{\tIII}\frac{N_1 N_2 }{(2\pi)^2 N_{12}} \big( \frac{(2\pi)^3 n_2 n_3 }{N_2 N_3}+\frac{(2\pi)^3 n_3 n_1}{N_3 N_1} \big)}]$.
\bea \label{eq:PhaseA1A2dA3}
\mathbf{Z} _0(\text{sym.twist}) =\exp[{ \frac{ -2\pi \, \ti \, p_{\tIII} n_3 }{ N_3} \frac{n_2 N_1+ n_1 N_2}{N_{12} }\big)}].\;\;\;\;\;\;\;\;
\eea
We can arbitrarily choose $n_1,n_2,n_3$ to determine the level identification of $p_{\tIII}$ from the flux identification.
The finest  level identification is determined from choosing the smallest $n_3$ and the smallest ${n_2 N_1+ n_1 N_2}$.
We choose $n_3=1$.
By Chinese remainder theorem, we can choose ${n_2 N_1+ n_1 N_2}=\gcd(N_1,N_2) \equiv N_{12}$.
Thus Eq.(\ref{eq:PhaseA1A2dA3}) yields
$\mathbf{Z} _0(\text{sym.twist}) =\exp[{ \frac{ -2\pi \, \ti \, p_{\tIII}}{ N_3} }]$. It is apparent that the flux identification implies the level identification
\be \label{eq:levelA1A2dA3_2}
p_{\tIII} \simeq p_{\tIII} +N_3.
\ee
Eq.(\ref{eq:levelA1A2dA3}),(\ref{eq:levelA1A2dA3_2}) and their linear combination together
imply the finest level $p_{\tIII}$ identification
\be \label{eq:levelA1A2dA3_N123}
p_{\tIII} \simeq p_{\tIII} +\gcd(N_{12},N_{3}) \simeq p_{\tIII} +N_{123}.
\ee
Overall, our derivation suggests that Eq.(\ref{eq:AppA1A2dA3}) has $p_{\tIII} \in \Z_{N_{123}}$.}

\subsubsection{ ${\int A_1 C_2}$} \label{sec:A1C2}

Similar to Sec.\ref{sec:A1B2}, we can introduce a flat 3-form $C$ field with $\dd C=0$ such that $\mathbf{Z} _0(\text{sym.twist})= \exp[{\;\ti\; k_{\tII}\int A_1 C_2}]$
can capture a similar physics of $\int A_1 A_2 \dd A_2$. The large gauge transformation and flux identification constrain respectively,
\bea
&&\frac{(2\pi) k_{\tII}}{N_u } = 0{\pmod{1}},\\
&& (2\pi) k_{\tII} \simeq (2\pi) k_{\tII} +N_1 N_2.
\eea
with $u \in \{ 1,2\}$. We derive
\bea
&&\mathbf{Z} _0(\text{sym.twist})
= \exp[{\;\ti\; p_{\tII}\frac{N_1 N_2}{(2\pi) N_{12}} \int_{\cM^{4}}  A_1 C_2}], \;\;\;\;
\eea
with $p_{\tII}$ defined by $p_{\tII} \pmod{N_{12}}$, thus $p_{\tII} \in \Z_{N_{12}}$.

\subsubsection{ ${\int A_1 A_2 B_3}$}

Similar to Sec.\ref{sec:A1B2}, \ref{sec:A1C2}, in 3+1D, by dimensional counting,
we can also introduce $\mathbf{Z} _0(\text{sym.twist})= \exp[{\;\ti\; k_{}\int A_1  A_2 B_3}]$.
The large gauge transformation and the flux identification yield
\bea
&&\frac{(2\pi)^2 k_{}}{N_u N_v } = 0{\pmod{1}},\\
&& (2\pi)^2 k_{} \simeq (2\pi)^2 k_{} +N_1 N_2 N_3.
\eea
We thus derive
\bea
&&\mathbf{Z} _0(\text{sym.twist})
= \exp[{\;\ti\; p_{\tIII}\frac{N_1 N_2 N_3}{(2\pi)^2 N_{123}} \int_{\cM^{4}}  A_1 A_2 B_3}],\;\;\;\;\;\;\;\;\;\;\;
\eea
with $p_{\tIII}$ defined by $p_{\tIII} \pmod{N_{123}}$ with $p_{\tIII} \in \Z_{N_{123}}$.

\subsection{Cases for Fermionic  SPTs} \label{sec:fermion}
Throughout the main text, we have been focusing on
the bosonic  SPTs, which elementary particle contents are all bosons.
Here we comment how the rules of fermionic  SPTs can be modified from bosonic  SPTs.
Due to that the fermionic particle is allowed, 
by exchanging two identical fermions will gain a fermionic statistics $\ep^{\ti \pi}=-1$, thus\\
{\bf $\bullet$ Large gauge transformation}: The $\mathbf{Z} _0$ invariance under the allowed large gauge transformation implies
the volume-integration must be invariant mod $\pi$ (instead of bosonic case with mod $2\pi$), because inserting a fermion into the system does not change the SPT class of system.
Generally, there are no obstacles to go through the analysis and level-quantization for fermions, except that we need to be careful about the flux identification.
Below we give an example of U(1) symmetry bosonic/fermionc  SPTs, and we will leave the details of other cases for future studies. 

\subsection{ { U$(1)^m$ symmetry} bosonic and fermionic  SPTs}

For U$(1)^m$ symmetry, one can naively generalize the above results from a viewpoint of $G=\Pi_m \Z_{N}= (\Z_{N}{})^m$ with $N \to \infty$.
This way of thinking is intuitive (though not mathematically rigorous), but guiding us to obtain U$(1)^m$ symmetry classification.
We find the classification is trivial for even $(d+1)$D, due to $F_u \wedge F_v \wedge \dots$
({where $F= \dd A$ is the field strength,} here $u,v$ can be either the same or different U$(1)$ gauge fields)
is only a surface term, not a bulk topological term.
For odd $(d+1)$D, we can define the lower type action:
$
\mathbf{Z} _0(\text{sym.twist})
= \exp[{\;\ti\; \frac{2\pi k}{(\frac{d+2}{2})!(2\pi)^{(d+2)/2}}  \int A_u \wedge F_v \wedge \dots}].
$
Meanwhile we emphasize that other type of
actions, such as the {\it top} type, $k\int A_1 A_2  \dots A_{d+1}$ form, or any other terms involve with more than one $A$
(e.g. $k\int A_{u_1} A_{u_2} \dots \dd A_{u.}$) will be trivial SPT class
for U$(1)^m$ case -
since its coefficient $k$ no longer stays finite for $N \to \infty$ of $(Z_{N})^m$ symmetry  SPTs, so the level $k$ is not well-defined.
For physically relevant $2+1$D,
$k \in 2\Z$ for bosonic  SPTs, $k \in \Z$ for fermionic  SPTs via Sec.\ref{sec:fermion}.
Thus, we will have a $\Z^m \times \Z^{m(m-1)/2}$ classification for U$(1)^m$ symmetry boson,
and the fermionic classification increases at least by shifting the bosonic $\Z \to 2\Z$. 
There may have even more extra classes by including Majorana boundary modes, 
which we will leave for future investigations.


\begin{widetext}
\end{widetext}
\onecolumngrid

\section{From ``Partition Functions of Fields'' to ``Cocycles of Group Cohomology'' and K\"unneth formula} \label{App:cocycles}

In Appendix \ref{App:levelQuant}, we have formulated the {\bf spacetime partition functions of probe fields} (e.g. $\mathbf{Z} _0(A(x))$, etc), which fields $A(x)$ take values at any coordinates $x$ on a continuous spacetime manifold $\cM$
with no dynamics.
On the other hand, it is known that, $(d+1)$D bosonic SPTs of symmetry group $G$ can be classified by the $(d+1)$-{th} cohomology group $\cH^{d+1}(G,\R/\Z)$\cite{Chen:2011pg}
(predicted to be complete at least for finite symmetry group $G$ without time reversal symmetry).
From this prediction that bosonic SPTs can be classified by group cohomology, 
our path integral on the discretized space lattice (or spacetime complex) shall be mapped to the {\bf partition functions of the cohomology group - the cocycles}.
In this section, we ask ``whether we can attain this correspondence from ``partition functions of fields'' to ``cocycles of group cohomology?''
Our answer is ``yes,''
we will bridge this beautiful correspondence between \emph{continuum field theoretic partition functions} and \emph{discrete cocycles} for any $(d+1)$D spacetime dimension for finite Abelian $G=\prod_u Z_{N_u}$.

\begin{center}
\begin{table}[!h]
\noindent
\makebox[\textwidth][c]{
\begin{tabular}{|c||c|c|c|}
\hline
(d+1)\text{dim}  & partition function $\mathbf{Z}$ &  $(d+1)$-cocycle $\omega_{d+1}$\\
 \hline\hline
0+1\tD&  $\exp(\ti \, p_{\tI} \int A_1) $ &  $ \exp \Big( \frac{2 \pi \ti p_{\tI}  }{N_{1}} \;  a_{1}\Big)$ \\[0mm]  \hline\hline
1+1\tD &  $\exp(\ti \,  p_{\tII}  \frac{N_1 N_2 }{(2\pi) N_{12}} \int A_1 A_2) $ & $ \exp \Big( \frac{2 \pi \ti p_{\tII}  }{N_{12}} \;  a_{1} b_{2}\Big)$  \\[0mm]  \hline\hline
2+1\tD &  $\exp(\ti \frac{p_{\tI} }{(2\pi) } \int A_1\dd A_1) $  &  $\exp \Big( \frac{2 \pi \ti {p_{\tI} }  }{N_{1}^{2}} \; a_{1}(b_{1} +c_{1} -[b_{1}+c_{1}]) \Big) $\\
            &  $\exp(\ti\, {p_{\tI} } \int C_1) $ (even/odd effect)  &  $\exp \Big( \frac{2 \pi \ti {p_{\tI} }  }{N_{1} } \, a_{1} b_{1} c_{1} \Big) $ \\ \hline
2+1\tD & $\exp(\ti \frac{p_{\tII} }{(2\pi) } \int A_1\dd A_2) $  & $\exp \Big( \frac{2 \pi \ti {p_{\tII} }  }{N_{1} N_{2}} \; a_{1}(b_{2} +c_{2} -[b_{2} +c_{2}]) \Big) $ \\
            & $\exp(\ti \,  p_{\tII} \frac{N_1 N_2 }{(2\pi) N_{12}}   \int A_1B_2) $ (even/odd effect) & $ \exp \Big( \frac{2 \pi \ti p_{\tII}   }{N_{12}} \;  a_{1} b_{2} c_{2}\Big)$   \\  \hline
2+1\tD & $\exp(\ti \, p_{\tIII}\frac{N_1 N_2 N_3}{(2\pi)^2 N_{123}}  \int A_1 A_2 A_3) $ & $ \exp \Big( \frac{2 \pi \ti p_{\tIII}  }{N_{123}} \;  a_{1}b_{2}c_{3} \Big)$ \\  \hline\hline
3+1\tD & $\exp(\ti \int p_{{ \tII(12)}}^{(1st)}\frac{N_1 N_2 }{(2\pi)^2 N_{12}}  A_1 A_2 \dd A_2) $  & ${\exp \big( \frac{2 \pi \ti p_{{ \tII(12)}}^{(1st)} }{ (N_{12} \cdot N_2  )   }    (a_1 b_2 )( c_2 +d_2 - [c_2+d_2  ]) \big)}$\\
            & $\exp(\ti \, p_{\tII} \frac{N_1 N_2 }{(2\pi) N_{12}}  \int A_1 C_2) $ (even/odd effect)  & ${\exp \big( \frac{2 \pi \ti p_{\tII} }{ N_{12}    }    a_1 b_2 c_2 d_2 \big)}$ \\  \hline
3+1\tD & $\exp(\ti \int p_{{ \tII(12)}}^{(2nd)}\frac{N_1 N_2 }{(2\pi)^2 N_{12}}  A_2 A_1 \dd A_1) $  & ${\exp \big( \frac{2 \pi \ti p_{{ \tII(12)}}^{(2nd)} }{ (N_{12} \cdot N_1  )   }    (a_2 b_1 )( c_1 +d_1 - [c_1+d_1  ]) \big)}$ \\
            & $\exp(\ti \, p_{\tII} \frac{N_1 N_2 }{(2\pi) N_{12}}  \int A_2 C_1) $ (even/odd effect)  & ${\exp \big( \frac{2 \pi \ti p_{\tII} }{ N_{12}    }    a_2 b_1 c_1 d_1 \big)}$ \\  \hline
3+1\tD & $\exp(\ti \, p_{{ \tIII(123)}}^{(1st)}\frac{N_1 N_2 }{(2\pi)^2 N_{12}}  \int (A_1 A_2) \dd A_3) $ &  ${\exp \big( \frac{2 \pi \ti p_{{ \tIII(123)}}^{(1st)} }{ (N_{12} \cdot N_3 )  }  (a_1 b_2 )( c_3 +d_3 - [c_3+d_3  ]) \big) }$\\
            & $\exp(\ti \, p_{\tIII}\frac{N_1 N_2 N_3}{(2\pi)^2 N_{123}}  \int A_1 A_2 B_3) $ (even/odd effect) &  ${\exp \big( \frac{2 \pi \ti p_{\tIII} }{ N_{123}   }  a_1 b_2 c_3 d_3 \big) }$\\  \hline
3+1\tD &  $\exp(\ti \, p_{{ \tIII(123)}}^{(2nd)}\frac{N_3 N_1 }{(2\pi)^2 N_{31}}  \int (A_3 A_1) \dd A_2 ) $ &  ${\exp \big( \frac{2 \pi \ti p_{{ \tIII(123)}}^{(2nd)} }{ (N_{31} \cdot N_2 )  }  (a_3 b_1 )( c_2 +d_2 - [c_2 +d_2  ]) \big) }$ \\
            & $\exp(\ti \, p_{\tIII}\frac{N_1 N_2 N_3}{(2\pi)^2 N_{123}}  \int A_3 A_1 B_2 ) $ (even/odd effect) &  ${\exp \big( \frac{2 \pi \ti p_{\tIII} }{ N_{123}   }  a_3 b_1 c_2 d_2 \big) }$\\  \hline
3+1\tD & $[\exp(\ti \, p_{\tIV}\frac{N_1 N_2 N_3 N_4 }{(2\pi)^3 N_{1234}}  \int A_1 A_2 A_3 A_4) ]$ & $\exp \big( \frac{2 \pi \ti p_{{ \tIV}}^{}}{ N_{1234} }  a_1 b_2 c_3 d_4 \big)$ \\  \hline\hline
4+1\tD &  $\exp(\ti \, \frac{p_{\tI} }{(2\pi)^2 }  \int A_1\dd A_1 \dd A_1) $  & ${\exp \left( \frac{2 \pi \ti {p_{\tI} } }{ (N_1)^3 } \;a_1( b_1+c_1 - [b_1+c_1 ]) ( d_1 +e_1 - [d_1 +e_1 ]) \right) }$ \\  \hline
4+1\tD & $\dots$ & $\dots$\\  \hline
4+1\tD & $\exp(\ti \, p_{\tV}\frac{N_1 N_2 N_3 N_4 N_5}{(2\pi)^4 N_{12345}}  \int A_1 A_2 A_3 A_4A_5) $ & $\exp \big( \frac{2 \pi \ti  p_{\tV} }{ N_{12345} }  a_1 b_2 c_3 d_4 e_5 \big)$\\  \hline
\hline
\end{tabular}
}\hspace*{0mm}
\caption{Some derived results on the correspondence between the {\bf spacetime partition function
of probe fields} (the second column) and  the {\bf cocycles of the cohomology group} (the third column) for any finite Abelian group $G=\prod_u Z_{N_u}$.
The even/odd effect means that whether their corresponding cocycles are nontrivial or trivial(as coboundary) depends on the level $p$ and $N$ (of the symmetry group $Z_N$) is even/odd.
Details are explained in Sec \ref{subsec:GCcocycle}.
}
\label{table:cocyclefact}
\end{table}
\end{center}

\twocolumngrid

\subsection{Correspondence}

The partition functions
in Appendix \ref{App:levelQuant} have been treated with careful proper level-quantizations via large gauge transformations and flux identifications.
For $G=\prod_u Z_{N_u}$, the field $A_u, B_u, C_u$, etc, take values in $Z_{N_u}$ variables, thus we can express them as
\be \label{eq:MapABC}
A_u \sim \frac{2\pi g_u}{N_u}, \;\, B_u \sim \frac{2\pi g_u h_u}{N_u}, \;\, C_u \sim \frac{2\pi g_u h_u l_u}{N_u}
\ee
with $g_u, h_u, l_u \in Z_{N_u}$. Here 1-form $A_u$ takes $g_u$ value on one link of a $(d+1)$-simplex, 2-form $B_u$ takes $g_u,h_u$ values on two different links and
3-form $C_u$ takes $g_u,h_u, l_u$ values on three different links of a $(d+1)$-simplex. These correspondence suffices for the flat probe fields.

In other cases, we also need to interpret the non-flat $\dd A\neq 0$ at the monodromy defect as the external inserted fluxes, thus we identify
\be \label{eq:MapdA}
\dd A_u \sim \frac{2\pi (g_u+h_u -[g_u+h_u] )}{N_u},
\ee
here $[g_u+h_u] \equiv g_u+h_u \pmod{N_u}$. Such identification ensures $\dd A_u$ is a multiple of $2\pi$ flux, therefore it is consistent with the constraint Eq.(\ref{eq:sumdA}) at the continuum limit.
Based on the Eq.(\ref{eq:MapABC})(\ref{eq:MapdA}), we derive the correspondence in Table \ref{table:cocyclefact},
from the continuum path integral $\mathbf{Z} _0(\text{sym.twist})$ of fields to a U(1) function as the discrete partition function.
In the next subsection, we will verify the U(1) functions in the last column in Table \ref{table:cocyclefact} indeed are the cocycles $\omega_{d+1}$ of cohomology group.
Such a correspondence has been explicitly pointed out in our previous work Ref.\onlinecite{Wang:2014oya} and applied to derive the cocycles.

\onecolumngrid

\begin{center}
\begin{table}[!h]
\noindent
\makebox[\textwidth][r]{
\begin{tabular}{|c||c|c|c|}
\hline
(d+1)\text{dim} & Partition function 
$\mathbf{Z}$  of ``fields''  & $p \in \cH^{d+1}(G,\R/\Z)$ & K\"unneth formula in $\cH^{d+1}(G,\R/\Z)$  \\
 \hline\hline
0+1\tD & $\exp(\ti \,p..\int A_1) $ & $\Z_{N_1}$ &$\cH^{1}(\Z_{N_1}, \R/\Z) $   \\[0mm]  \hline \hline
1+1\tD & $\exp(\ti \,p..\int A_1 A_2) $ & $\Z_{N_{12}}$ &$ \cH^{1}(\Z_{N_1}, \R/\Z) \boxtimes_\Z \cH^{1}(\Z_{N_2}, \R/\Z) $\\[0mm]  \hline \hline
2+1\tD & $\exp(\ti \,p..\int A_1\dd A_1) $  & $\Z_{N_{1}}$  &$ \cH^{3}(\Z_{N_1}, \R/\Z)$    \\ \hline
2+1\tD & $\exp(\ti \,p..\int A_1\dd A_2) $  & $\Z_{N_{12}}$ &$\cH^{1}(\Z_{N_1}, \R/\Z) \otimes_\Z \cH^{1}(\Z_{N_2}, \R/\Z)$  \\  \hline
2+1\tD & $\exp(\ti \,p..\int A_1 A_2 A_3) $ & $\Z_{N_{123}}$ &$[\cH^1(\Z_{N_1},\R/\Z)\boxtimes_\Z \cH^1(\Z_{N_2},\R/\Z)]\boxtimes_\Z \cH^1(\Z_{N_3}, \R/\Z)$  \\  \hline \hline
3+1\tD & $\exp(\ti \,p..\int A_1 A_2 \dd A_2) $ & $\Z_{N_{12}}$ &$ \cH^{1}(\Z_{N_1}, \R/\Z) \boxtimes_\Z \cH^{3}(\Z_{N_2}, \R/\Z)  $  \\  \hline
3+1\tD & $\exp(\ti \,p..\int A_2 A_1 \dd A_1) $ & $\Z_{N_{12}}$ &$ \cH^{1}(\Z_{N_2}, \R/\Z) \boxtimes_\Z \cH^{3}(\Z_{N_1}, \R/\Z)  $  \\  \hline
3+1\tD & $\exp(\ti \,p..\int (A_1 A_2) \dd A_3) $ & $\Z_{N_{123}}$ &$ {  [ \cH^{1}(\Z_{N_1} , \R/\Z) \boxtimes_\Z \cH^{1}(\Z_{N_2}, \R/\Z) ]     \otimes_\Z  \cH^1(\Z_{N_3}, \R/\Z) }   $  \\  \hline
3+1\tD & $\exp(\ti \,p..\int (A_1 \dd A_2) A_3) $  &  $\Z_{N_{123}}$ &$  [\cH^{1}(\Z_{N_1} , \R/\Z) \otimes_\Z \cH^{1}(\Z_{N_2},\R/\Z) ] \boxtimes_\Z  \cH^1(\Z_{N_3}, \R/\Z) ) $ \\  \hline
3+1\tD & $\exp(\ti \,p..\int A_1 A_2 A_3 A_4) $ &$\Z_{N_{1234}}$ &$  \big[[\cH^1(\Z_{N_1}, \R/\Z)  \boxtimes_\Z \cH^1(\Z_{N_2}, \R/\Z)]  \boxtimes_\Z  \cH^1(\Z_{N_3}, \R/\Z) \big] \boxtimes_\Z  \cH^1(\Z_{N_4}, \R/\Z)$  \\  \hline \hline
4+1\tD & $\exp(\ti \,p..\int A_1\dd A_1 \dd A_1) $ &$\Z_{N_{1}}$ &$\cH^{5}(\Z_{N_1},  \R/\Z)$   \\  \hline
4+1\tD & $\exp(\ti \,p..\int A_1 \dd A_1 \dd A_2) $  &$\Z_{N_{12}}$ &$\cH^{3}(\Z_{N_1}  \R/\Z) \otimes_\Z \cH^{1}(\Z_{N_2}  ,\R/\Z)$  \\  \hline
4+1\tD & $\exp(\ti \,p..\int A_2 \dd A_2 \dd A_1) $ &$\Z_{N_{12}}$ &$\cH^{3}(\Z_{N_2}  ,\R/\Z) \otimes_\Z \cH^{1}(\Z_{N_1}  ,\R/\Z)$   \\  \hline
4+1\tD & $\exp(\ti \,p..\int A_1 \dd A_1 A_2 A_3 ) $  & $\Z_{N_{123}}$ & ${ \big[  [\cH^{3}(\Z_{N_1}, \R/\Z)  \boxtimes_\Z \cH^{1}(\Z_{N_2}, \R/\Z) ]   \boxtimes_\Z  \cH^1( \Z_{N_3}, \R/\Z) \big]   }$ \\  \hline
4+1\tD & $\exp(\ti \,p..\int A_2 \dd A_2 A_1 A_3) $ & $\Z_{N_{123}}$ & ${ \big[  [\cH^{3}(\Z_{N_2}, \R/\Z)  \boxtimes_\Z \cH^{1}(\Z_{N_1}, \R/\Z) ]   \boxtimes_\Z  \cH^1( \Z_{N_3}, \R/\Z) \big]   }$  \\  \hline
4+1\tD & $\exp(\ti \,p..\int A_1 \dd A_2 \dd A_3) $ & $\Z_{N_{123}}$ & ${ { [\cH^1( \Z_{N_1}, \R/\Z) \otimes_Z \cH^1( \Z_{N_2}, \R/\Z) ]\otimes_Z \cH^1( \Z_{N_3}, \R/\Z)  } }$  \\  \hline
4+1\tD & $\exp(\ti \,p..\int A_1 A_2 A_3 \dd A_3) $ & $\Z_{N_{123}}$ & ${ \big[  [\cH^{1}(\Z_{N_1}, \R/\Z)  \boxtimes_\Z \cH^{1}(\Z_{N_2},\R/\Z) ]   \boxtimes_\Z  \cH^3( \Z_{N_3}, \R/\Z) \big]   }$  \\  \hline
4+1\tD & $\exp(\ti \,p..\int  A_1 \dd A_2 A_3 A_4) $ & $\Z_{N_{1234}}$ & ${ \Big[ \big[  [\cH^{1}(\Z_{N_1}, \R/\Z)  \otimes_\Z \cH^{1}(\Z_{N_2}, \R/\Z) ]   \boxtimes_\Z  \cH^1( \Z_{N_3}, \R/\Z) \big]
\boxtimes_\Z  \cH^1( \Z_{N_4}, \R/\Z) \Big]   }$  \\  \hline
4+1\tD & $\exp(\ti \,p..\int  A_1 A_2 \dd A_3 A_4) $  & $\Z_{N_{1234}}$ & ${ \Big[ \big[  [\cH^{1}(\Z_{N_1}, \R/\Z)  \boxtimes_\Z \cH^{1}(\Z_{N_2}, \R/\Z) ]   \otimes_\Z  \cH^1( \Z_{N_3}, \R/\Z) \big]
\boxtimes_\Z  \cH^1( \Z_{N_4}, \R/\Z) \Big]   }$ \\  \hline
4+1\tD & $\exp(\ti \,p..\int  A_1 A_2 A_3 \dd A_4) $  & $\Z_{N_{1234}}$ & ${ \Big[ \big[  [\cH^{1}(\Z_{N_1}, \R/\Z)  \boxtimes_\Z \cH^{1}(\Z_{N_2}, \R/\Z) ]   \boxtimes_\Z  \cH^1( \Z_{N_3}, \R/\Z) \big]
\otimes_\Z  \cH^1( \Z_{N_4},\R/\Z) \Big]   }$ \\  \hline
4+1\tD & $\exp(\ti \,p.. \int A_1 A_2 A_3 A_4A_5) $ &$\Z_{N_{12345}}$ &$  \Big[\big[[\cH^1(\Z_{N_1})  \boxtimes_\Z \cH^1(\Z_{N_2})]  \boxtimes_\Z  \cH^1(\Z_{N_3}) \big] \boxtimes_\Z  \cH^1(\Z_{N_4})
\Big] \boxtimes_\Z  \cH^1(\Z_{N_5}) $  \\  \hline
\end{tabular}
}\hspace*{-12mm}
\caption{ {\bf From partition functions of fields to K\"unneth formula}. 
Here we consider a finite Abelian group $G=\prod_u Z_{N_u}$.
The field theory result can map to the derived facts about the cohomology group and its cocycles.
Here the level-quantization is shown in a shorthand way with only $p ..$ written, the explicit coefficients can be found in Table \ref{table:cocycleclass}.
In some row, we abbreviate $\cH^1(\Z_{n_j},\R/\Z) \equiv \cH^1(\Z_{n_j})$.
The torsion product $\text{Tor}_1^\Z \equiv \boxtimes_\Z $ evokes a wedge product  $\wedge$ structure in the corresponding field theory,
while the tensor product $\otimes_\Z$ evokes appending an extra exterior derivative $\wedge  \dd$ structure in the corresponding field theory.
This simple observation maps the field theoretic path integral to its correspondence in K\"unneth formula.
}
\label{table:cocycleclass}
\end{table}
\end{center}

\begin{table}[!h]
\begin{tabular}{|c||c|c|c|c|c|c|c|c|c|}
\hline
      &Type I& Type II & Type III &Type  IV &Type  V  &Type  VI & $\dots$ & $\dots$ & \\[0mm]  \hline
      &$\Z_{N_i}$& $\Z_{N_{ij}}$& $\Z_{N_{ijl}}$ &  $\Z_{N_{ijlm}}$ &  $\Z_{{\gcd} \otimes^5_{i}(N^{(i)})}$
      &  $\Z_{{\gcd} \otimes^6_{i}(N_i)}$ & $\Z_{{\gcd} \otimes^m_{i}(N_i)}$ & $\Z_{{\gcd} \otimes^{d-1}_{i}N_i}$ & $\Z_{{\gcd} \otimes^d_{i}N^{(i)}}$\\[0mm]  \hline
$\cH^1(G,\R/\Z)$ &$1$& $$& $$ & $$ & $$ & $$ & $$ & $$ & $$ \\[0mm]  \hline
$\cH^2(G,\R/\Z)$&$0$  &$1$ & $$ & $$ & $$ & $$ & $$ & $$ & $$\\[0mm]  \hline
$\cH^3(G,\R/\Z)$ &$1$ & $1$ & $1$ & $$ & $$ & $$ & $$ & $$ & $$\\ \hline
$\cH^4(G,\R/\Z)$ &$0$ & $2$ & $2$ & $1$ & $$ & $$ & $$ & $$ & $$\\ \hline
$\cH^5(G,\R/\Z)$ & $1$ & $2$ & $4$ & $3$ & $1$ & $$ & $$ & $$ & $$\\ \hline
$\cH^6(G,\R/\Z)$ & $0$ & $3$ & $6$ & $7$ & $4$ & $1$ & $$ & $$ & $$\\ \hline
$\cH^d(G,\R/\Z)$ & $\frac{(1-(-1)^d)}{2}$ & $\frac{d}{2}-\frac{(1-(-1)^d)}{4}$ & $\dots$ & $\dots$ & $\dots$ & $\dots$ & $\dots$ & $d-2$ & $1$\\ \hline
\end{tabular}
\caption{
The table shows the exponent of the $\Z_{{\gcd} \otimes^m_{i}(N_i)}$ class in the cohomology group
$\cH^d(G,\R/\Z)$ for a finite Abelian group $G=\underset{u=1}{\prod^k} Z_{N_u}$.
Here we define a shorthand of $\Z_{{\gcd} (N_i,N_j)}\equiv
\Z_{N_{ij}} \equiv  \Z_{{\gcd} \otimes^2_{i}(N_i)}$, etc also for other higher
gcd.  Our definition of the Type $m$  is from its number ($m$) of cyclic gauge
groups in the gcd class $\Z_{{\gcd} \otimes^m_{i}(N_i)}$.  The number of
exponents can be systematically obtained by adding all the numbers of the
previous column from the top row to a row before the wish-to-determine number.
This table in principle can be independently derived by gathering the data of
Table \ref{table:cocycleclass} from field theory approach.
For example, we can derive
$\mathcal{H}^5(G,\R/\Z) =  \underset{1 \leq i < j < l<m <n \leq k}{\prod} \Z_{N_i} \times (\Z_{N_{ij}})^2\times (\Z_{N_{ijl}})^4 \times (\Z_{N_{ijlm}})^3  \times  \Z_{N_{ijlmn}} $, etc.
Thus, we can use field theory to \emph{derive} the group cohomology result.
}
\label{table:Hgroup}
\end{table}

We remark that the field theoretic path integral's level $p$ quantization and its mod relation also provide an independent way (apart from group cohomology)
to count the number of types of partition functions for a given symmetry group $G$ and a given spacetime dimension.
Such the modular $p$ is organized in (the third column of) Table \ref{table:cocycleclass}.
In addition,
one can further deduce the {\bf K\"unneth formula}(the last column of Table \ref{table:cocycleclass}) from a field theoretic partition function viewpoint.
Overall, this correspondence from field theory can be an
independent powerful tool to {\it derive}  the group cohomology and extract the classification data (such as Table \ref{table:Hgroup}).

\subsection{Cohomology group and cocycle conditions} \label{subsec:GCcocycle}
To verify that the last column of Table \ref{table:cocyclefact} (bridged from the field theoretic partition function) 
are indeed cocycles of a cohomology group,
here we briefly review the cohomology group $\mathcal{H}^{d+1}(G,\R/\Z)$ (equivalently as $\mathcal{H}^{d+1}(G,\tU(1))$ by $\R/\Z=\tU(1)$),
which is the ${(d+1)}$th-cohomology group of G over G module U(1).
Each class in $\mathcal{H}^{d+1}(G,\R/\Z)$ corresponds to a distinct $(d+1)$-cocycles.
The $n$-cocycles is a $n$-cochain, in addition
they satisfy the $n$-cocycle-conditions $\delta \omega=1$.
The $n$-cochain is a mapping of $\omega_{}^{}(a_1,a_2,\dots,a_n)$:  $G^n \to \tU(1)$ (which inputs
$a_i \in G$, $i=1,\dots, n$, and outputs a $\tU(1)$ value).
The $n$-cochain satisfies the group multiplication rule:
\be
(\omega_{1}\cdot\omega_{2})(a_1,\dots,a_n)= \omega_{1}^{}(a_1,\dots,a_n)\cdot \omega_{2}^{}(a_1,\dots,a_n),
\ee
thus form a group.
The coboundary operator $\delta$
\bea \label{eq:delta}
&&\delta \sfc(g_1, g_2, \dots,  g_{n+1}) \equiv
\sfc(g_2, \dots,  g_{n+1})  \sfc(g_1, \dots,  g_{n})^{(-1)^{n+1}}
\cdot \prod_{j=1}^{n} \sfc(g_1, \dots, \; g_j g_{j+1},\; \dots,  g_{n+1})^{(-1)^{j}},
\eea
\twocolumngrid
\noindent
which defines the $n$-cocycle-condition $\delta \omega=1$.
The $n$-cochain forms a group $\text{C}^n$,
while the $n$-cocycle forms its subgroup $\text{Z}^n$.
The distinct $n$-cocycles are not equivalent via $n$-coboundaries,
where Eq.(\ref{eq:delta}) also defines the $n$-coboundary relation:
if n-cocycle $\omega_n$ can be written as
$\omega_n = \delta \Omega_{n-1}$, for any $(n-1)$-cochain $\Omega_{n+1}$, then we
say this $\omega_n$ is a $n$-coboundary.
Due to $ \delta^2=1$, thus we know that the $n$-coboundary further forms a subgroup $\text{B}^n$ . In short,
$
\text{B}^n \subset \text{Z}^n \subset \text{C}^n
$
The $n$-cohomology group is precisely a kernel $\text{Z}^n$ (the group of $n$-cocycles) mod out image $\text{B}^n$ (the group of $n$-coboundary) relation:
\bea
\cH^n(G,\R/\Z)= \text{Z}^n /\text{B}^n.
\eea
For other details about group cohomology (especially Borel group cohomology here), we suggest to read Ref.\onlinecite{Chen:2011pg,{Wang:2014oya},{deWildPropitius:1996gt}} and Reference therein.

To be more specific cocycle conditions, for finite Abelian group $G$, the 3-cocycle condition for 2+1D is (a pentagon relation),
\be
\delta \omega(a,b,c,d)=\frac{ \omega(b,c,d) \omega(a,bc,d) \omega(a,b,c) }{\omega(ab,c,d)\omega(a,b,cd) }=1
\ee

The 4-cocycle condition for 3+1D is
\be
\delta \omega(a,b,c,d,e)=\frac{ \omega(b,c,d,e) \omega(a,bc,d,e) \omega(a,b,c,de) }{\omega(ab,c,d,e)\omega(a,b,cd,e) \omega(a,b,c,d) }=1
\ee
The 5-cocycle condition for 4+1D is
\bea
&&\delta \omega(a,b,c,d,e,f)=\frac{ \omega(b,c,d,e,f) \omega(a,bc,d,e,f) }{\omega(ab,c,d,e,f) } \nonumber \\
&&\cdot \frac{  \omega(a,b,c,de,f) \omega(a,b,c,d,e)  }{ \omega(a,b,cd,e,f) \omega(a,b,c,d,ef)   }=1
\eea
We verify that the U(1) functions (mapped from a field theory derivation) in the last column of Table \ref{table:cocyclefact} indeed satisfy cocycle conditions.
Moreover, those  {\bf partition functions purely involve with 1-form $A$ or its field-strength (curvature) $\dd A$ are strictly cocycles but not coboundaries}.
These imply that those terms with only $A$ or $\dd A$ are the precisely nontrivial cocycles in the cohomology group for classification.

However, we find that {\bf partition functions  involve with 2-form $B$, 3-form $C$ or higher forms, although are cocycles but sometimes may also be coboundaries}
at certain quantized level $p$ value. For instance,
for those cocycles correspond to the partition functions of 
${p} \int C_1$,
$p_{} \frac{N_1 N_2 }{(2\pi) N_{12}}   \int A_1B_2$,
$p_{} \frac{N_1 N_2 }{(2\pi) N_{12}}  \int A_1 C_2$,
$p_{} \frac{N_1 N_2 }{(2\pi) N_{12}}  \int A_2 C_1$,
$p_{}\frac{N_1 N_2 N_3}{(2\pi)^2 N_{123}}  \int A_1 A_2 B_3$,
$p_{}\frac{N_1 N_2 N_3}{(2\pi)^2 N_{123}}  \int A_3 A_1 B_2$, etc (which involve with higher forms $B$, $C$),
we find that for $G=(Z_2)^n$ symmetry, $p=1$ are in the nontrivial class (namely not a coboundary),
$G=(Z_4)^n$ symmetry, $p=1, 3$ are in the nontrivial class (namely not a coboundary).
However, for $G=(Z_3)^n$ symmetry of all $p$ and $G=(Z_4)^n$ symmetry at $p=2$,
 are in the trivial class (namely a coboundary), etc.
This indicates an {\bf even-odd effect}, sometimes these cocycles are nontrivial, but sometimes are trivial as coboundary,
depending on the level $p$ is even/odd and the symmetry group $(Z_N)^n$ whether $N$ is even/odd.
{\bf Such an even/odd effect also bring complication into the validity of nontrivial cocycles, thus this is another reason that
we study only field theory involves with only 1-form $A$ or its field strength $\dd A$.
The cocycles composed from $A$ and $\dd A$ in Table \ref{table:cocyclefact} are always nontrivial and are not coboundaries.}

We finally point out that the concept of {\bf boundary term in field theory} (the surface or total derivative term) is connected to the
concept of  {\bf coboundary in the cohomology group}.
For example, $\int (\dd A_1) A_2 A_3$ are identified as the {\bf coboundary} of the linear combination of $\int A_1 A_2 (\dd A_3)$ and $\int A_1 (\dd A_2) A_3$.
Thus, by counting the number of distinct field theoretic actions (not identified by {\bf boundary term})
is precisely counting the number of distinct field theoretic actions (not identified by  {\bf coboundary}).
Such an observation matches the field theory classification to the group cohomology classification shown in Table \ref{table:Hgroup}.
Furthermore, we can map the field theory result to the K\"unneth formula listed in Table \ref{table:cocycleclass}, via the correspondence:
\bea
\int A_1   &\sim& \cH^{1}(\Z_{N_1},\R/\Z)\\
\int A_1 \dd A_1 &\sim& \cH^{3}(\Z_{N_1},\R/\Z)\\
\int A_1 \dd A_1 \dd A_1 &\sim& \cH^{5}(\Z_{N_1},\R/\Z)\\
\text{Tor}_1^\Z \equiv \boxtimes_\Z  &\sim& \wedge \\
\otimes_\Z  &\sim& \wedge \dd 
\eea
\bea
\int A_1 \wedge A_2  &\sim& \cH^{1}(\Z_{N_1}, \R/\Z) \boxtimes_\Z \cH^{1}(\Z_{N_2}, \R/\Z)\;\;\;\;\; \;\;\;\;\\
\int A_1 \wedge \dd A_2  &\sim& \cH^{1}(\Z_{N_1}, \R/\Z) \otimes_\Z \cH^{1}(\Z_{N_2}, \R/\Z) \;\;\;\;\; \;\;\;\;\\
 &\dots& \nonumber
\eea

To summarize, in this section, we show that, at lease for finite Abelian symmetry group $G=\prod^k_{i=1} Z_{N_i}$,
field theory can be systematically formulated, via the level-quantization developed in Appendix \ref{App:levelQuant},
we can count the number of classes of  SPTs. Explicit examples are organized
in Table \ref{table:cocyclefact}, \ref{table:cocycleclass}, \ref{table:Hgroup}, where we show that our field theory approach can exhaust all bosonic SPT classes (at least as complete as) in group cohomology:
\bea
&& \cH^2(G,\R/\Z) = \prod_{1 \leq i < j  \leq k}  \Z_{N_{ij}} \label{H2}\\
&& \cH^3(G,\R/\Z) = \prod_{1 \leq i < j < l \leq k}  \Z_{N_i} \times \Z_{N_{ij}} \times  \Z_{N_{ijl}} \label{H3}\\
&&\mathcal{H}^4(G,\R/\Z) =  \prod_{1 \leq i < j < l<m \leq k}
         (\Z_{N_{ij}})^2
         \times (\Z_{N_{ijl}})^2
         \times \Z_{N_{ijlm}}\;\;\; \;\; \;\;\;\;\label{eq:H4}\\
&& \;\; \;\;\;\;\;\; \;\;\;\;\;\; \;\;\;\; \dots   \nonumber
\eea
and we also had addressed the correspondence between field theory and K\"unneth formula.

\onecolumngrid

\section{SPT Invariants, Physical Observables and Dimensional Reduction} \label{sec:app-SPTprobe}

In this section, we comment more about the SPT invariants from probe field partition functions, and the derivation of SPT Invariants
from dimensional reduction, using both a continuous field theory approach and a discrete cocycle approach. We focus on finite Abelian $G=\prod_u Z_{N_u}$ bosonic  SPTs.

First, recall from the main text using a continuous field theory approach, we can summarize the dimensional reduction as a diagram below:
\begin{equation}
\xymatrix@R=2mm@C=6mm{
1+1\text{D} & 2+1\text{D}       & 3+1\text{D}       & \cdots       & d+1\text{D} \\
A_1 A_2  & A_1 A_2 A_3 \ar[l]&  A_1 A_2 A_3 A_4 \ar[l] & \cdots \ar[l] & A_1 A_2 \dots A_{d+1}\ar[l]\\
&&&&\\
    & A_v \dd A_w & A_u A_v \dd A_w \ar[uull] \ar[l] & \dots \ar[l] &
}
\end{equation}
There are basically (at least) two ways for dimensional reduction procedure:\\
$\bullet (i)$ One way is the left arrow $\leftarrow$ procedure,
which compactifies one spatial direction $x_u$ as a $S^1$ circle while a gauge field $A_u$ along that $x_u$ direction takes $Z_{N_u}$ value by $\oint_{S^1} A_u=2\pi n_u/N_u$.\\
$\bullet (ii)$ Another way of dimensional reduction is the up-left arrow $\nwarrow$,  where the space is designed as $M^2\times M^{d-2}$, where
a 2-dimensional surface $M^2$ is drilled with holes or punctures of monodromy defects with $\dd A_w$ flux, via $ \Ointint \sum \dd A_w = 2 \pi n_w$ under the condition
Eq.(\ref{eq:sumdA}). As long as the net flux through all the holes is not zero ($n_w \neq 0$), the dimensionally reduced partition functions can be nontrivial SPTs at lower dimensions.
We summarize their physical probes in Table \ref{table:probeDR} and in its caption.
\begin{center}
\begin{table}[!h]
\noindent
\makebox[\textwidth][l]{
\begin{tabular}{l l }
  \hline\\[-1.5mm]
  \textbf{Physical Observables} & Dimensional reduction of SPT invariants and probe-feild actions\\
  \hline\\[-0.5mm]
 $\bullet$ {\bf degenerate zero energy modes}\cite{Wang:2014tia} of 1+1D SPT &   \;\;\;\;\;$A_1 A_2  \leftarrow A_1 A_2 A_3 \leftarrow  A_1 A_2 A_3 A_4 \leftarrow \cdots \;\;\; $ \\
   (projective representation of $Z_{N_1} \times Z_{N_2}$ symmetry) &   \;\;\;\;\;$A_1 A_2  \leftarrow  A_u A_v  \dd A_w \leftarrow \cdots$ \\[2mm]
    \hline\\[-1.5mm]
  $\bullet$ edge modes on monodromy defects of 2+1D SPT - gapless, &  \;\;\;\;\;$A_v  \dd A_w  \leftarrow  A_u A_v  \dd A_w \leftarrow \cdots \;\;\; $       \\
  \;\; or gapped with {\bf induced fractional quantum numbers}\cite{Wang:2014tia} &  \;\;\;\;\;       \\
    $\bullet$ {\bf braiding statistics of monodromy defects}\cite{{LG1209},{Chenggu},{WL1437},{Wang:2014oya}} &  \;\;\;\;\;       \\[0.5mm]
  \hline
\end{tabular}
}\hspace*{2mm}
\caption{ We discuss two kinds of dimensional-reducing outcomes and their physical observables. The first kind reduces to $\int A_1 A_2$ type action of 1+1D  SPTs, where
its 0D boundary modes carries a projective representation of the remained symmetry $Z_{N_1} \times Z_{N_2}$, due to its action is a nontrivial element of $\cH^2(Z_{N_1} \times Z_{N_2},\R/\Z)$.
This projective representation also implies the degenerate zero energy modes near the 0D boundary.
The second kind reduces to $\int A_v \dd A_w$ type action of 2+1D  SPTs, where its physical observables are either gapless edge modes at the monodromy defects, or
gapped edge by symmetry-breaking domain wall which induces fractional quantum numbers. One can also detect this SPTs by its nontrivial braiding statistics of gapped monodromy defects
(particles/strings in 2D/3D for $\int A\dd A$\;/\;$\int AA\dd A$ type actions).
}
\label{table:probeDR}
\end{table}
\end{center}

Second, we can also apply a discrete cocycle approach (to verify the above field theory result). We only need to use the slant product,
which sends a $n$-cochain $\sfc$ to a $(n-1)$-cochain $i_g \sfc$:
\bea
i_g \sfc(g_1, g_2, \dots,  g_{n-1}) \equiv  \sfc(g, g_1, g_2,\dots,g_{n-1})^{(-1)^{n-1}} \cdot \prod_{j=1}^{n-1} \sfc(g_1,\dots,g_j, (g_1\dots g_j)^{-1} \cdot g \cdot  (g_1\dots g_j),\dots,g_{n-1})^{(-1)^{n-1+j}}, \;\;\;\;
\eea
\twocolumngrid
\noindent
with $g_i \in G$. Let us consider Abelian group  $G$, in 2+1D, where we dimensionally reduce by sending a $3$-cocycle to a $2$-cocycle:
\bea
\sfC_a(b,c) \equiv i_a\omega(b,c) =\frac{\omega(a,b,c) \omega(b,c,a) }{ \omega(b,a,c) }.
\eea
In 3+1D, we dimensionally reduce by sending a $4$-cocycle to a $3$-cocycle:
\bea
\sfC_a(b,c,d) \equiv i_a \omega(b,c,d) =\frac{\omega(b,a,c,d) \omega(b,c,d,a) }{ \omega(a,b,c,d) \omega(b,c,a,d) }. \;\;\;
\eea
These dimensionally-reduced cocycles from Table \ref{table:cocyclefact}'s last column would agree with the field theory dimensional reduction structure and its predicted SPT invariants.

\twocolumngrid

\bibliography{}

\begin{thebibliography}{32}
\expandafter\ifx\csname natexlab\endcsname\relax\def\natexlab#1{#1}\fi
\expandafter\ifx\csname bibnamefont\endcsname\relax
  \def\bibnamefont#1{#1}\fi
\expandafter\ifx\csname bibfnamefont\endcsname\relax
  \def\bibfnamefont#1{#1}\fi
\expandafter\ifx\csname citenamefont\endcsname\relax
  \def\citenamefont#1{#1}\fi
\expandafter\ifx\csname url\endcsname\relax
  \def\url#1{\texttt{#1}}\fi
\expandafter\ifx\csname urlprefix\endcsname\relax\def\urlprefix{URL }\fi
\providecommand{\bibinfo}[2]{#2}
\providecommand{\eprint}[2][]{\url{#2}}


\bibitem [{\citenamefont {Ginzburg}\ and\ \citenamefont
  {Landau}(1950)}]{GL5064}%
  \BibitemOpen
  \bibfield  {author} {\bibinfo {author} {\bibfnamefont {V.~L.}\ \bibnamefont
  {Ginzburg}}\ and\ \bibinfo {author} {\bibfnamefont {L.~D.}\ \bibnamefont
  {Landau}},\ }  
   {\bibfield  {journal} {\bibinfo  {journal} {Zh.
  Eksp. Teor. Fiz.}\ }\textbf {\bibinfo {volume} {20}},\ \bibinfo {pages}
  {1064} (\bibinfo {year} {1950})} 
  
  
\bibitem [{\citenamefont {Landau}\ and\ \citenamefont
  {Lifschitz}(1958)}]{LanL58}%
  \BibitemOpen
  \bibfield  {author} {\bibinfo {author} {\bibfnamefont {L.~D.}\ \bibnamefont
  {Landau}}\ and\ \bibinfo {author} {\bibfnamefont {E.~M.}\ \bibnamefont
  {Lifschitz}},\ } 
  {\emph {\bibinfo {title} {Statistical Physics -
  Course of Theoretical Physics Vol 5}}}\ (\bibinfo  {publisher} {Pergamon},\
  \bibinfo {address} {London},\ \bibinfo {year} {1958}) 



 \bibitem[{\citenamefont{Wen}(1989)}]{Wtop}
\bibinfo{author}{\bibfnamefont{X.-G.} \bibnamefont{Wen}},
  \bibinfo{journal}{Phys. Rev. B} \textbf{\bibinfo{volume}{40}},
  \bibinfo{pages}{7387} (\bibinfo{year}{1989}).

\bibitem[{\citenamefont{Wen and Niu}(1990)}]{WNtop}
\bibinfo{author}{\bibfnamefont{X.-G.} \bibnamefont{Wen}} \bibnamefont{and}
  \bibinfo{author}{\bibfnamefont{Q.}~\bibnamefont{Niu}},
  \bibinfo{journal}{Phys. Rev. B} \textbf{\bibinfo{volume}{41}},
  \bibinfo{pages}{9377} (\bibinfo{year}{1990}).

\bibitem[{\citenamefont{Wen}(1990)}]{Wrig}
\bibinfo{author}{\bibfnamefont{X.-G.} \bibnamefont{Wen}},
  \bibinfo{journal}{Int. J. Mod. Phys. B} \textbf{\bibinfo{volume}{4}},
  \bibinfo{pages}{239} (\bibinfo{year}{1990}).
  

\bibitem{Chen:2011pg} 
  X.~Chen, Z.~-C.~Gu, Z.~-X.~Liu and X.~-G.~Wen,
  Phys.\ Rev.\ B {\bf 87}, 155114 (2013)
  [arXiv:1106.4772 [cond-mat.str-el]].
  

\bibitem[{\citenamefont{{Chen} et~al.}(2013)\citenamefont{{Chen}, {Gu}, {Liu},
  and {Wen}}}]{2013arXiv1301.0861C}
\bibinfo{author}{\bibfnamefont{X.}~\bibnamefont{{Chen}}},
  \bibinfo{author}{\bibfnamefont{Z.-C.} \bibnamefont{{Gu}}},
  \bibinfo{author}{\bibfnamefont{Z.-X.} \bibnamefont{{Liu}}}, \bibnamefont{and}
  \bibinfo{author}{\bibfnamefont{X.-G.} \bibnamefont{{Wen}}},
  \bibinfo{journal}{Science}  
  \textbf{\bibinfo{volume}{338}},
    \bibinfo{pages}{1604}
  (\bibinfo{year}{2012}),

\bibitem{Gu:2009dr} 
  Z.~-C.~Gu and X.~-G.~Wen,
  Phys.\ Rev.\ B {\bf 80}, 155131 (2009)
  [arXiv:0903.1069 [cond-mat.str-el]].


\bibitem{Pollmann:2009} 
F.~Pollmann, A.M.~Turner, E.~Berg, M.~Oshikawa
  Phys.\ Rev.\ B {\bf 81}, 064439 (2010)

    
  






\bibitem{Qi:2008ew} 
  X.~L.~Qi, T.~Hughes and S.~C.~Zhang,
  Phys.\ Rev.\ B {\bf 78}, 195424 (2008)
  
\bibitem{Wen:2013oza} 
  X.-G.~Wen,
  Phys.\ Rev.\ D {\bf 88}, no. 4, 045013 (2013)
  [arXiv:1303.1803 [hep-th]].
  
\bibitem{RyuZhang}
S.~Ryu and S.~-C.~Zhang, 
Phys.\ Rev.\ B {\bf 85}, 245132 (2012).
  
\bibitem{Wang:2013yta} 
  J.~Wang and X.~-G.~Wen,
  arXiv:1307.7480 [hep-lat].


 
\bibitem{Kapustin:2014lwa} 
  A.~Kapustin and R.~Thorngren,
  arXiv:1403.0617 [hep-th].



\bibitem{Kapustin:2014zva} 
  A.~Kapustin and R.~Thorngren,
  arXiv:1404.3230 [hep-th].
   
  
%
\bibitem{Vishwanath:2012tq} 
  A.~Vishwanath and T.~Senthil,
  Phys.\ Rev.\ X {\bf 3}, 011016 (2013).
  
\bibitem{K1467} 
  A.~Kapustin,
  arXiv:1403.1467 [cond-mat.str-el].
  
\bibitem{K1459} 
A.~Kapustin,
  arXiv:1404.6659 [cond-mat.str-el].
  
\bibitem{Freed:2014eja} 
  D.~S.~Freed,
  arXiv:1406.7278 [cond-mat.str-el].
  


\bibitem{Fidkowski:2013jua} 
  L.~Fidkowski, X.~Chen and A.~Vishwanath,
  Phys.\ Rev.\ X {\bf 3}, 041016 (2013)

\bibitem{CWang1} 
C. Wang and T. Senthil,
Phys.\ Rev.\ B {\bf 87}, 235122 (2013)

\bibitem{CWang2} 
C. Wang, A.~C.~Potter, and T. Senthil,  
Science {\bf 343}, 6171 (2014)

%

	
\bibitem{Metlitski:2013uqa} 
  M.~A.~Metlitski, C.~L.~Kane and M.~P.~A.~Fisher,
  Phys.\ Rev.\ B {\bf 88}, no. 3, 035131 (2013)

\bibitem[Chen et al.(2014)]{2014arXiv1403.6491C} X.~Chen, F.~J. Burnell, 
A.~Vishwanath,  L.~Fidkowski, 
arXiv:1403.6491 



  
\bibitem{Sule:2013qla} 
  O.~M.~Sule, X.~Chen and S.~Ryu,
  Phys.\ Rev.\ B {\bf 88}, 075125 (2013)
  [arXiv:1305.0700 [cond-mat.str-el]].
  
\bibitem{Wang:2014tia} 
  J.~Wang, L.~H.~Santos, X.~-G.~Wen,
  arXiv:1403.5256 [cond-mat.str-el].
  
\bibitem{Cho:2014jfa} 
  G.~Y.~Cho, J.~C.~Teo and S.~Ryu,
  arXiv:1403.2018 [cond-mat.str-el].
  
\bibitem{Hsieh:2014lba} 
  C.~-T.~Hsieh, O.~M.~Sule, G.~Y.~Cho, S.~Ryu and R.~Leigh,
  arXiv:1403.6902 [cond-mat.str-el].

\bibitem{Kravec:2013pua} 
  S.~M.~Kravec and J.~McGreevy,
  Phys.\ Rev.\ Lett.\  {\bf 111}, 161603 (2013)
  
  

\bibitem{KW14}
  L.~Kong and X.~-G.~Wen,
  arXiv:1405.5858 



\bibitem{Chen:2010gda} 
  X.~Chen, Z.~C.~Gu and X.~G.~Wen,
  Phys.\ Rev.\ B {\bf 82}, 155138 (2010)
  [arXiv:1004.3835 [cond-mat.str-el]].

\bibitem[{\citenamefont{Verstraete et~al.}(2005)\citenamefont{Verstraete,
  Cirac, Latorre, Rico, and Wolf}}]{VCL0501}
\bibinfo{author}{\bibfnamefont{F.}~\bibnamefont{Verstraete}},
  \bibinfo{author}{\bibfnamefont{J.~I.} \bibnamefont{Cirac}},
  \bibinfo{author}{\bibfnamefont{J.~I.} \bibnamefont{Latorre}},
  \bibinfo{author}{\bibfnamefont{E.}~\bibnamefont{Rico}}, \bibnamefont{and}
  \bibinfo{author}{\bibfnamefont{M.~M.} \bibnamefont{Wolf}},
  \bibinfo{journal}{Phys. Rev. Lett.} \textbf{\bibinfo{volume}{94}},
  \bibinfo{pages}{140601} (\bibinfo{year}{2005}).

\bibitem[{\citenamefont{Vidal}(2007)}]{V0705}
\bibinfo{author}{\bibfnamefont{G.}~\bibnamefont{Vidal}},
  \bibinfo{journal}{Phys. Rev. Lett.} \textbf{\bibinfo{volume}{99}},
  \bibinfo{pages}{220405} (\bibinfo{year}{2007}).
  

%
\bibitem{H8364}%
  \BibitemOpen
  \bibfield{author}{%
  \bibinfo {author} {\bibfnamefont{F.~D.~M.}\ \bibnamefont{Haldane}},\ }%
  \bibfield{journal}{%
  \bibinfo {journal} {Physics Letters A}\ }%
  \textbf{\bibinfo {volume} {93}},\ \bibinfo {pages} {464} (\bibinfo {year}
  {1983})%

%
\bibitem{AKL8877}%
  \BibitemOpen
  \bibfield{author}{%
  \bibinfo {author} {\bibfnamefont{I.}~\bibnamefont{Affleck}}, \bibinfo
  {author} {\bibfnamefont{T.}~\bibnamefont{Kennedy}}, \bibinfo {author}
  {\bibfnamefont{E.~H.}\ \bibnamefont{Lieb}},\ and\ \bibinfo {author}
  {\bibfnamefont{H.}~\bibnamefont{Tasaki}},\ }%
  \bibfield{journal}{%
  \bibinfo {journal} {Commun. Math. Phys.}\ }%
  \textbf{\bibinfo {volume} {115}},\ \bibinfo {pages} {477} (\bibinfo {year}
  {1988})%

%
\bibitem{TI4} M. Z. Hasan, C. L. Kane, Rev. Mod. Phys. \textbf{82}, 3045
(2010). 
%
%
\bibitem{TI6} X.-L. Qi, S.-C. Zhang, Rev. Mod. Phys. \textbf{83}, 1057 (2011).

\bibitem{TI5} J. E. Moore, Nature \textbf{464}, 194 (2010).



\bibitem{Schwarz:1978cn} 
  A.~S.~Schwarz,
  Lett.\ Math.\ Phys.\  {\bf 2}, 247 (1978).

\bibitem{Witten:1988hf} 
  E.~Witten,
  Commun.\ Math.\ Phys.\  {\bf 121}, 351 (1989).
  
\bibitem{Witten:1988ze} 
  E.~Witten,
  Commun.\ Math.\ Phys.\  {\bf 117}, 353 (1988).
  
  
  

\bibitem{gaugesym}
Since gauge symmetry is not a real symmetry but only a redundancy,
we can use gauge symmetry to describe the topological order which has no real global symmetry.



\bibitem{Dijkgraaf:1989pz} 
  R.~Dijkgraaf and E.~Witten,
  Commun.\ Math.\ Phys.\  {\bf 129}, 393 (1990).

\bibitem{dim}
Overall we denote $(d+1)$D as d dimensional space and one dimensional time, and $d$D for d dimensional space.



\bibitem{Levin:2009} 
  M.~Levin and A.~Stern,
  Phys.\ Rev.\ Lett. {\bf 103}, 196803 (2009)
  [arXiv:0906.2769 [cond-mat.str-el]].

  
  
\bibitem{Levin:2012ta} 
  M.~Levin and A.~Stern,
  Phys.\ Rev.\ B {\bf 86}, 115131 (2012)
  [arXiv:1205.1244 [cond-mat.str-el]].
  
  

\bibitem{Lu:2012dt} 
  Y.~-M.~Lu and A.~Vishwanath,
  Phys.\ Rev.\ B {\bf 86}, 125119 (2012)
  [arXiv:1205.3156 [cond-mat.str-el]].
  
\bibitem{Chenggu} Meng Cheng and Zheng-Cheng Gu, Phys. Rev. Lett. 112, 141602 (2014) 

\bibitem{Ye:2013upa} 
  P.~Ye and J.~Wang,
  Phys.\ Rev.\ B {\bf 88}, 235109 (2013)
  [arXiv:1306.3695 [cond-mat.str-el]].

\bibitem{LiuWen}
Z.-X. Liu, X.-G. Wen,
Phys. Rev. Lett. 110, 067205 (2013)

\bibitem{Bi:2013oza} 
  Z.~Bi, A.~Rasmussen and C.~Xu,
  arXiv:1309.0515 [cond-mat.str-el].



\bibitem{refAppendix}
A systematic step-by-step derivation and many more examples 
can be found in the Supplemental Material.
\cblue{In Appendix A, we provide more details on the derivation of SPTs partition functions of fields with level quantization.
In Appendix B, we provide the correspondence between SPTs' ``partition functions of fields'' to ``cocycles of group cohomology.''
In Appendix C, we systematically organize SPT invariants and their physical observables by dimensional reduction.}


\bibitem{Wilczek:1984dh} 
  F.~Wilczek and A.~Zee,
  Phys.\ Rev.\ Lett.\  {\bf 52}, 2111 (1984).
  
\bibitem[{\citenamefont{Keski-Vakkuri and Wen}(1993)}]{KW9327}
\bibinfo{author}{\bibfnamefont{E.}~\bibnamefont{Keski-Vakkuri}}
  \bibnamefont{and} \bibinfo{author}{\bibfnamefont{X.-G.} \bibnamefont{Wen}},
  \bibinfo{journal}{Int. J. Mod. Phys. B} \textbf{\bibinfo{volume}{7}},
  \bibinfo{pages}{4227} (\bibinfo{year}{1993}).
  
\bibitem{Zhang:2011jd} 
  Y.~Zhang, T.~Grover, A.~Turner, M.~Oshikawa and A.~Vishwanath,
  Phys.\ Rev.\ B {\bf 85}, 235151 (2012)

\bibitem[Moradi 
\& Wen(2014)]{2014arXiv1401.0518M}  H. Moradi,  X.-G.~Wen, arXiv:1401.0518 


\bibitem[{\citenamefont{Moradi and Wen}(2014)}]{MW14}
\bibinfo{author}{\bibfnamefont{H.}~\bibnamefont{Moradi}} \bibnamefont{and}
  \bibinfo{author}{\bibfnamefont{X.-G.} \bibnamefont{Wen}},
  \bibinfo{journal}{arXiv:1404.4618}. 



  
  
\bibitem{nAbgauge}
Here the geometric phase or Berry phase has a gauge structure.
Note that the non-Abelian gauge structure of degenerate ground states can appear even for Abelian topological order with Abelian braiding statistics.


\bibitem{Wen:2013ue} 
  X.~-G.~Wen,
  Phys.\ Rev.\ B {\bf 89}, 035147 (2014)
  [arXiv:1301.7675 [cond-mat.str-el]].
  
  
\bibitem{Hung:2013cda} 
  L.~-Y.~Hung and X.~-G.~Wen,
  Phys.\ Rev.\ B {\bf 89}, 075121 (2014)
  [arXiv:1311.5539 [cond-mat.str-el]].


\bibitem{Hung:2013qpa} 
  L.~-Y.~Hung and Y.~Wan,
  arXiv:1308.4673 [cond-mat.str-el].

\bibitem[{\citenamefont{Levin and Gu}(2012)}]{LG1209}
\bibinfo{author}{\bibfnamefont{M.}~\bibnamefont{Levin}} \bibnamefont{and}
  \bibinfo{author}{\bibfnamefont{Z.-C.} \bibnamefont{Gu}},
  \bibinfo{journal}{Phys. Rev. B} \textbf{\bibinfo{volume}{86}},
  \bibinfo{pages}{115109} (\bibinfo{year}{2012}), 
  
  
  \bibitem[{\citenamefont{Gu and Levin}(2014)}]{LG14}
    \bibinfo{author}{\bibfnamefont{Z.-C.} \bibnamefont{Gu}},
\bibnamefont{and}
\bibinfo{author}{\bibfnamefont{M.}~\bibnamefont{Levin}} 
  \bibinfo{journal}{Phys. Rev. B} \textbf{\bibinfo{volume}{89}},
  \bibinfo{pages}{201113(R)} (\bibinfo{year}{2014}), 

\bibitem{Barkeshli:2013yta} 
  M.~Barkeshli, C.~-M.~Jian and X.~-L.~Qi,
  Phys.\ Rev.\ B {\bf 88}, 235103 (2013)
  [arXiv:1305.7203 [cond-mat.str-el]].


\bibitem{Santos:2013uda} 
  L.~H.~Santos and J.~Wang,
  Phys.\ Rev.\ B {\bf 89}, 195122 (2014)
  [arXiv:1310.8291 [quant-ph]].
  
  
\bibitem[{\citenamefont{Wang and Levin}(2014)}]{WL1437}
\bibinfo{author}{\bibfnamefont{C.}~\bibnamefont{Wang}} \bibnamefont{and}
  \bibinfo{author}{\bibfnamefont{M.}~\bibnamefont{Levin}}
  (\bibinfo{year}{2014}), \eprint{arXiv:1403.7437}.
  
\bibitem[{\citenamefont{Jiang et~al.}(2014)\citenamefont{Jiang, Mesaros, and
  Ran}}]{JMR1462}
\bibinfo{author}{\bibfnamefont{S.}~\bibnamefont{Jiang}},
  \bibinfo{author}{\bibfnamefont{A.}~\bibnamefont{Mesaros}}, \bibnamefont{and}
  \bibinfo{author}{\bibfnamefont{Y.}~\bibnamefont{Ran}} (\bibinfo{year}{2014}),
  \eprint{arXiv:1404.1062}.

\bibitem{Wang:2014oya} 
  J.~Wang and X.~-G.~Wen,
  arXiv:1404.7854 [cond-mat.str-el].

\bibitem{Jian:2014vfa} 
  C.~-M.~Jian and X.~-L.~Qi,
  arXiv:1405.6688 [cond-mat.str-el].
  

 
 
 
    





\bibitem{deWildPropitius:1996gt} 
  M.~de Wild Propitius,
  Nucl.\ Phys.\ B {\bf 489}, 297 (1997)

\bibitem{Kapustin:2014gua} 
  A.~Kapustin and N.~Seiberg,
  JHEP {\bf 1404}, 001 (2014)

\bibitem{Levin_talk}
This term has also been noticed by
M.~Levin at Princeton PCTS talk on \emph{Braiding statistics and symmetry-protected topological phases} (2014).


\bibitem{Chen:201301}
X.~Chen, Y.~-M.~Lu, A.~Vishwanath,
Nature Communications 5, 3507, 2014
 [arXiv:1303.4301 [cond-mat.str-el]].
 
\bibitem{Lu:2013wna} 
  Y.~-M.~Lu and D.~-H.~Lee,
  Phys.\ Rev.\ B {\bf 89}, 205117 (2014)
  

  
 %
\bibitem{Jackiw:1975fn} 
  R.~Jackiw and C.~Rebbi,
  Phys.\ Rev.\ D {\bf 13}, 3398 (1976).

 %
\bibitem{Goldstone:1981kk} 
  J.~Goldstone and F.~Wilczek,
  Phys.\ Rev.\ Lett.\  {\bf 47}, 986 (1981).
    
   

\bibitem{E8}
The E$_8$ quantum Hall state of 2+1D has a perturbative gravitational anomaly on its 1+1D boundary
via the gravitational Chern-Simons 3-form.


\bibitem{Kane_Fisher}
C. L. Kane and M. P. A. Fisher, Phys. Rev. Lett. {\bf 76}, 3192
(1996)

    
\bibitem{Wen:2014zga} 
  X.~G.~Wen,
  Phys.\ Rev.\ B {\bf 91}, 205101 (2015)
  doi:10.1103/PhysRevB.91.205101
  [arXiv:1410.8477 [cond-mat.str-el]].
  
    
      
\end{thebibliography}

\end{document}